\documentclass[aps,prl,superscriptaddress,notitlepage,nopacs,amsmath,amstex,amssymb,citeautoscript,longbibliography,floatfix,letter,twocolumn]{revtex4-2}
\usepackage{graphicx}
\usepackage{subfigure}
\usepackage{adjustbox}
\usepackage{bm}
\usepackage{color}
\usepackage{braket}
\usepackage{standalone}
\usepackage{multirow}
\usepackage{tikz}
\usepackage{mathrsfs}
\usepackage{dsfont}
\usepackage{comment}
\usepackage{amsmath}
\usepackage[colorlinks,bookmarks=true,citecolor=blue,linkcolor=red,urlcolor=blue]{hyperref}
\usepackage{cleveref}
\usepackage{orcidlink}
\renewcommand{\vec}[1]{\mbox{\boldmath$#1$}}

\newcommand{\addACB}[1]{{\color{orange}{#1}}}

\graphicspath{{}}

\begin{document}

\title{Microscopic Model for Fractional Quantum Hall Nematics} 

\author{Songyang Pu}
\thanks{These authors have contributed equally to this work.}
\affiliation{School of Physics and Astronomy, University of Leeds, Leeds LS2 9JT, United Kingdom}
\affiliation{Department of Physics and Astronomy, The University of Tennessee, Knoxville, TN 37996, USA}

\author{Ajit C. Balram\orcidlink{0000-0002-8087-6015}}
\thanks{These authors have contributed equally to this work.}
\affiliation{Institute of Mathematical Sciences, CIT Campus, Chennai 600113, India}
\affiliation{Homi Bhabha National Institute, Training School Complex, Anushaktinagar, Mumbai 400094, India} 

\author{Joseph Taylor\orcidlink{0009-0005-9793-9501}}
\affiliation{School of Physics and Astronomy, University of Leeds, Leeds LS2 9JT, United Kingdom}

\author{Eduardo Fradkin\orcidlink{0000-0001-6837-463X}}
\email{efradkin@illinois.edu}
\affiliation{Department of Physics, University of Illinois at Urbana-Champaign,
1110 West Green Street, Urbana, Illinois 61801, USA}
\affiliation{Anthony J. Leggett Institute for Condensed Matter Theory,
University of Illinois at Urbana-Champaign, 1110 West Green Street, Urbana, Illinois 61801, USA}

\author{Zlatko Papi\'c\orcidlink{0000-0002-8451-2235}}
\email{z.papic@leeds.ac.uk}
\affiliation{School of Physics and Astronomy, University of Leeds, Leeds LS2 9JT, United Kingdom}

\date{\today}

\begin{abstract}
Geometric fluctuations of the density mode in a fractional quantum Hall (FQH) state can give rise to a nematic FQH phase, a topological state with a spontaneously broken rotational symmetry. While experiments on FQH states in the second Landau level have reported signatures of putative FQH nematics in anisotropic transport, a realistic model for this state has been lacking. We show that the standard model of particles in the lowest Landau level interacting via the Coulomb potential realizes the FQH nematic transition, which is reached by a progressive reduction of the strength of the shortest-range Haldane pseudopotential. Using exact diagonalization and variational wave functions, we demonstrate that the FQH nematic transition occurs when the system's neutral gap closes in the long-wavelength limit while the charge gap remains open. We confirm the symmetry-breaking nature of the transition by demonstrating the existence of a ``circular moat'' potential in the manifold of states with broken rotational symmetry, while its geometric character is revealed through the strong fluctuations of the nematic susceptibility and Hall viscosity. 
\end{abstract}

\maketitle

\textbf{\textit{Introduction.}} Nematicity in the fractional quantum Hall (FQH) effect provides an intriguing link between the notions of topology and spontaneous symmetry breaking~\cite{Fradkin2010}. The FQH nematic (FQHN) phase~\cite{Mulligan10, Maciejko13, You14}  has a finite charge gap that leads to a quantized plateau in the Hall resistance, reminiscent of incompressible FQH fluids~\cite{Tsui82}. However, while the latter are also gapped to \textit{neutral} excitations, such as a pair of fundamental quasielectron and quasihole, the neutral gap in the FQHN phase is expected to vanish as a consequence of the spontaneous breaking of continuous rotational symmetry. Symmetry breaking occurs in many QH systems with the prominent examples being skyrmions~\cite{Sondhi93, Barrett95, Schmeller95, Balram15d} and charge-density-wave (CDW) phases, such as stripes, bubbles, and Wigner crystals~\cite{Koulakov96, Lam84, Levesque84, Moessner96, Balents96, Williams91, Engel97, Du99, Lilly99, Cooper99a, Xia04, Zhu10, Baer15, Deng16, Ma20, Schreiber20}. Unlike these, the FQHN maintains translation invariance but only breaks rotational symmetry about the $z$-axis perpendicular to the two-dimensional electron gas (2DEG) \cite{Fradkin99}.
The breaking of a continuous symmetry distinguishes the FQHN from other examples of discrete nematics~\cite{Abanin2010, Parameswaran2019} in multivalley materials~\cite{Shayegan2006, Eng2007, Feldman16}, frustrated magnets~\cite{Huang22}, and moir\'e superlattices~\cite{Jiang2019, Cao2021, RubioVerdu2022}.

\begin{figure}[b]
    \centering
    \includegraphics[width=0.99\linewidth]{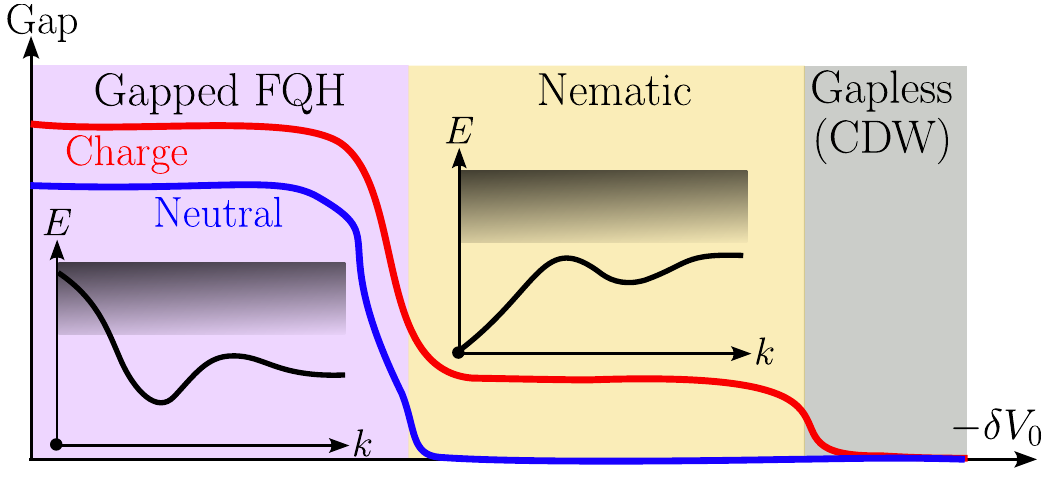}
    \caption{Schematic of the phase diagram and the FQHN transition for bosons, driven by varying the shortest-range potential from its Coulomb value by $\delta V_0$. In the nematic phase, the charge gap remains open, while the neutral gap has closed due to the presence of a Goldstone mode associated with the spontaneous breaking of continuous rotation symmetry. Upon further softening of the shortest-range repulsion, the system becomes fully gapless, e.g., by also breaking translation symmetry (CDW). 
    }
    \label{fig: sketch}
\end{figure}

A general mechanism believed to give rise to a FQHN from a proximate incompressible FQH state is the softening of the magnetoroton mode in the long-wavelength limit \cite{You14, Maciejko13},  Fig.~\ref{fig: sketch}. The magnetoroton mode is a collective density wave excitation~\cite{Pinczuk93, Kang01, Kukushkin09}  that occurs in many FQH states, including the Laughlin~\cite{Laughlin83} and Moore-Read~\cite{Moore91} states. At long wavelengths, the mode is described by the Girvin, MacDonald, and Platzman (GMP) ansatz~\cite{Girvin85, Girvin86}. Unlike the topological properties of FQH states, such as the Hall conductance, the long-wavelength limit of the GMP mode has a \textit{geometric} character: in an effective field theory, it behaves as a quadrupole that can be described by a quantum metric~\cite{Haldane11, Can14, Ferrari2014, Gromov14, Bradlyn15, You16, Gromov17, Nguyen18}.

Experiments in the second LL with tilted magnetic fields~\cite{Xia11} and hydrostatic pressure~\cite{Samkharadze15, Schreiber18} have observed anisotropic resistance that was attributed to nematicity. These experiments indeed suggest that nematic order can be proximate to incompressible FQH states or even \textit{coexist} with them. In addition, a Raman scattering experiment at filling factors $\nu{=}5/2$ and $7/2$ has provided evidence of an interplay between nematicity and electron pairing~\cite{Du19}. A major challenge in bridging the gap between these experimental observations and effective theories of FQHNs has been the lack of a microscopic model. Early work~\cite{Musaelian96} introduced a class of variational Laughlin-like wave functions that break the continuous rotational symmetry. More recently, Ref.~\cite{Regnault17} proposed a toy model exhibiting some signatures of the FQHN. For a similar class of short-range interacting models, Ref.~\cite{Yang20Nematic} formulated conditions for the quadrupole excitation to become gapless. However, the identification of FQHNs in realistic models 
has been lacking.

In this paper, we consider a quintessential model of the FQH effect -- Coulomb interaction projected to the lowest LL (LLL)~\cite{Prange87, Haldane85a, Rezayi88}. Contrary to the general belief that the softening of the shortest range component of the Coulomb interaction results in a destruction of the FQH liquid and direct transition to a CDW, we argue that this model realizes the scenario of the FQHN transition as shown in Fig.~\ref{fig: sketch}. 
Using exact diagonalization (ED), we demonstrate the hallmarks of the FQHN in this model: the GMP mode goes soft at long wavelengths and the neutral gap closes, while the charge gap remains open. Furthermore, we demonstrate the symmetry-breaking nature of the transition and show that the nematic susceptibility and the Hall viscosity both diverge in its vicinity. The results of exact numerics are supported by estimates of the gaps in larger systems using the GMP ansatz and composite fermion (CF) theory~\cite{Jain97}. Below we present illustrative results for the Laughlin state of bosons at filling factor $\nu_{b}{=}1/2$, which is relevant for recent experiments in cold atom setups~\cite{Leonard23}. However, in the Supplemental Material (SM)~\cite{SM}, we demonstrate that our results equally apply to the $\nu{=}1/3$ Laughlin state of electrons. 

\textbf{\textit{Model.}} We consider a standard model for the FQH effect where $N$ electrons, interacting pairwise via the Coulomb potential, are confined to a spherical surface~\cite{Haldane83}, with a Dirac monopole at the center, emanating a radial magnetic flux of strength $2Qhc/e$. The radius of the sphere is $R{=}\sqrt{Q}\ell_B$, where $\ell_B{=}\sqrt{\hbar c/eB}$ is the magnetic length. The Hamiltonian is translation- and rotation-invariant, which implies that the total orbital angular momentum $L$ and its $z$-component $M$ are good quantum numbers. The magnitude of the planar wavevector $k$ is given by $k{=}L/R$. The rotational invariance of the interaction implies that it can be expressed in terms of Haldane pseudopotentials $\{ V_m \}$, where $m{=}0,1,2,3,{\ldots}$ is the relative angular momentum of any pair of particles~\cite{Haldane83, Prange87}. Hence, $m$ is constrained by the statistics of the particles: for bosons, only even pseudopotentials $V_0$, $V_2$, $V_4$, etc. are relevant. 

We will consider the filling factor $\nu_b{=}1/2$ for bosons, at which the Laughlin state~\cite{Laughlin83} is realized if we set the magnetic monopole flux to $2Q{=}2N{-}\mathcal{S}$, with $\mathcal{S}{=}2$ denoting the Wen-Zee shift~\cite{Wen92}.  Starting from the Coulomb interaction, we soften the shortest-range component of the interaction potential and therefore the model is parametrized by a single number $\delta V_0{=}V_{0}{-}V_0^\mathrm{C}$, where $V_0^\mathrm{C}$ is the Coulomb interaction's $V_{0}$ value~\cite{Fano86}. 

\begin{figure}[tb]
    \centering    
    \includegraphics[width=\linewidth]{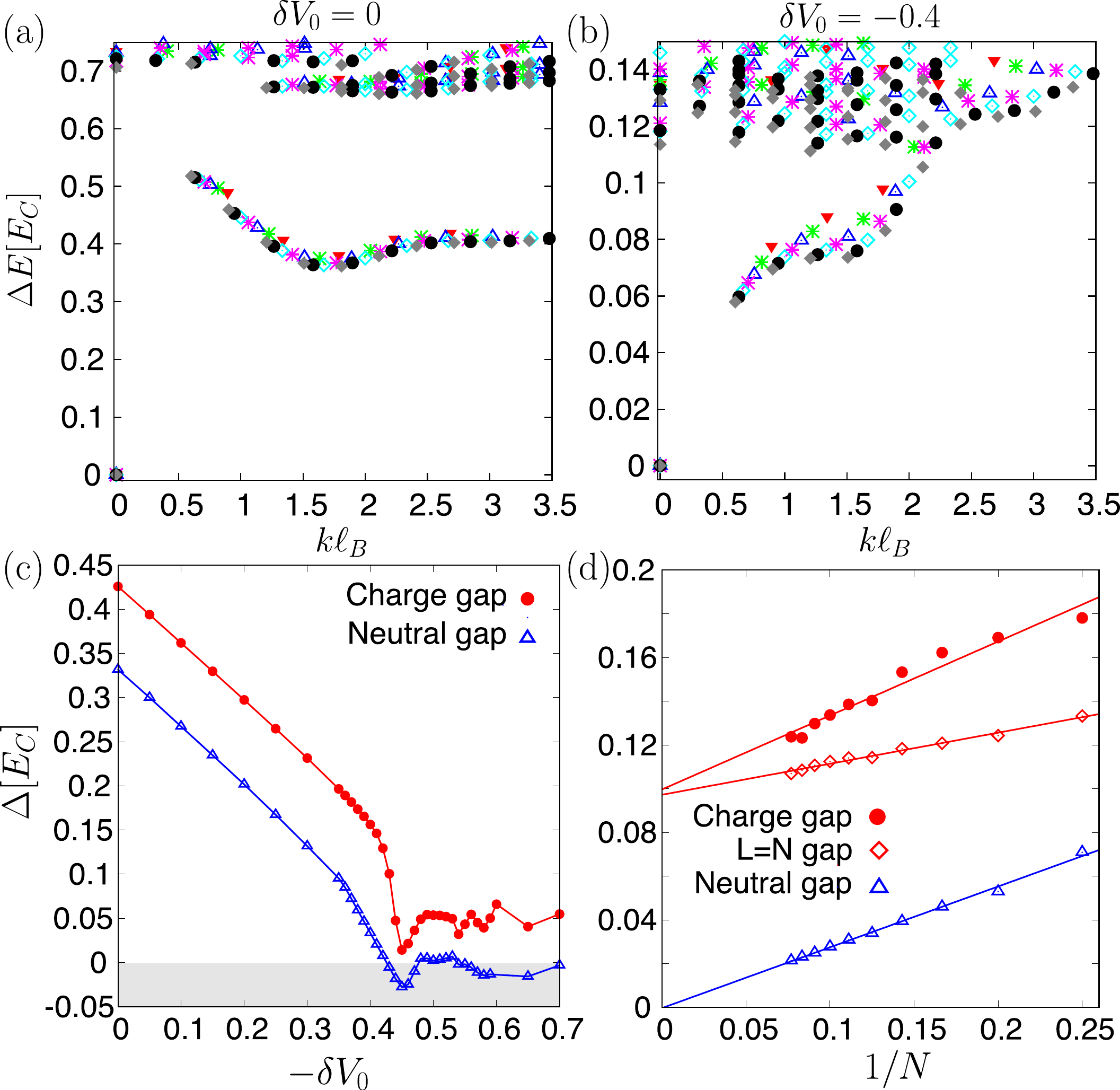}
    \caption{
    (a)-(b): Energy spectrum for bosons at $\nu_{b}{=}1/2$ in the LLL, with pure Coulomb potential (a) and softened by adding $\delta V_0 {=} {-}0.4$ pseudopotential (b), which pushes the system close to a critical point. Data is for system sizes $N{=}6{-}12$. All energies include the rescaling of the magnetic length, see SM~\cite{SM}. (c) Charge and neutral gaps (see text for the definition) of the bosonic $\nu_{b}{=}1/2$ state as a function of $\delta V_0$. (d) The scaling of the gaps as a function of $1/N$ for the fixed value $\delta V_0{=}{-}0.43$.
    }
    \label{fig: bosons_spectrum_gaps}
\end{figure}

\textbf{\textit{Spectral properties.}} We first analyze the energy spectra of the $\nu_{b}{=}1/2$ bosonic state obtained using ED, Fig.~\ref{fig: bosons_spectrum_gaps}(a)-(b). Throughout, energies are expressed in units of Coulomb energy, $E_C {\equiv} e^2/\epsilon \ell_B$. While there is a gapped magnetoroton mode for the pure Coulomb interaction [Fig.~\ref{fig: bosons_spectrum_gaps}(a)], the dispersion of the collective mode is significantly altered by reducing the $V_0$ component of the interaction. As we add a delta interaction of strength $\delta V_0 {=} {-}0.4$, the wave number corresponding to the roton minimum changes from $k\ell_B {\sim} 1.5$ to $k{\rightarrow}0$ [Fig.~\ref{fig: bosons_spectrum_gaps}(b)]. The softening of the magnetoroton mode implies that the neutral gap decreases as $V_{0}$ is reduced and potentially closes in the long-wavelength limit.

In Fig.~\ref{fig: bosons_spectrum_gaps}(c)-(d) we show the gaps as a function of $\delta V_0$.  We evaluate two types of gaps for accessible system sizes 
and perform their extrapolation to the thermodynamic limit in $1/N$: (i) the neutral gap is defined as the difference in the lowest two energies at the ground state flux, $2Q{=}2N{-}2$; (ii) the charge or transport gap, i.e., the energy to create an individual quasihole (occurs at flux $2Q{+}1$) and quasiparticle excitation (occurs at flux $2Q{-}1$). 
An alternative way to estimate the charge gap 
is to extrapolate the gap $\Delta^{L{=}N}$, the energy difference between the ground state and the lowest-lying state with orbital angular momentum $L{=}N$, at the ground state flux. The $L{=}N$ roton state is formed by a quasiparticle and a quasihole on opposite poles of the sphere. The interaction between these localized excitations vanishes in the thermodynamic limit, hence the charge gap is simply the sum of their individual energies.

The results in Fig.~\ref{fig: bosons_spectrum_gaps}(c) show a sharp drop in both the charge and neutral gaps as the interaction is softened by $\delta V_0 {\approx} {-}0.4$. More precisely, at $\delta V_0 {=} {-}0.43$, the neutral gap drops to zero (or slightly below, due to the uncertainty of the extrapolation), however, the charge gap, while significantly reduced compared to the pure Coulomb point, remains non-zero. This is demonstrated by the raw scaling data for gaps as a function of $1/N$ in Fig.~\ref{fig: bosons_spectrum_gaps}(d).  The reliability of the extrapolation is confirmed by independently calculating $\Delta^{L{=}N}$, which extrapolates to the same value as the charge gap, Fig.~\ref{fig: bosons_spectrum_gaps}(d).

\begin{figure}[bt]
    \centering    
    \includegraphics[width=0.95\linewidth]{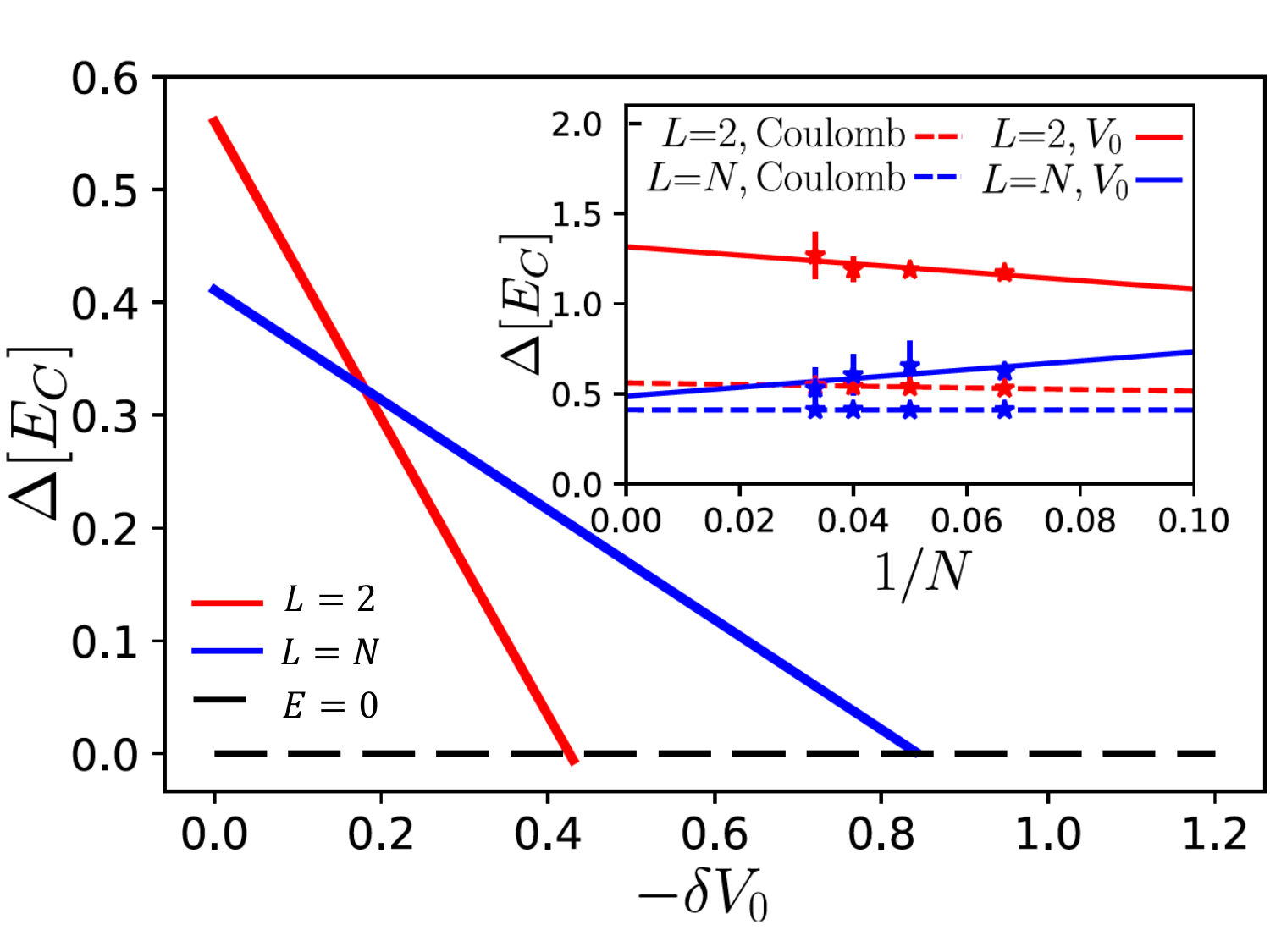}
    \caption{
     Variational estimate of the neutral gap (i.e., the gap of the $L{=}2$ GMP state) and charge gap (i.e., the gap of $L{=}N$ CF exciton state) as a function of $\delta V_0$ for bosons at $\nu_{b}{=}1/2$. The inset shows the finite-size extrapolation of the Coulomb and $V_0$ gaps of $L{=}2$ and $L{=}N$ trial states.
    }
    \label{fig: bosons_variational}
\end{figure}

\textbf{\textit{Variational wave functions.}} To access system sizes beyond ED, we employ GMP \cite{Girvin85} and composite fermion (CF)~\cite{Jain89} ansatz wave functions. Fig.~\ref{fig: bosons_variational} shows the gap estimates from such wave functions. We approximate the ground state by the bosonic Laughlin wave function~\cite{Laughlin83} which has a high overlap (upwards of 0.85 for all $N{\leq}14$) with the exact ground state across the range  $\delta V_{0}{\in}[{-}0.43,0]$. The exact spectra, Fig.~\ref{fig: bosons_spectrum_gaps}(b), show that the gap can be upper-bounded by the GMP excitation that carries 
angular momentum $L{=}2$ on the sphere~\cite{Pu23}. On the other hand, to estimate the charge gap, we use the CF exciton wave functions, which accurately describe the entire magnetoroton branch~\cite{Kamilla96b, Balram21d}. The CF exciton wave function of the Laughlin state in the long wavelength limit is identical to the GMP ansatz. We estimate the charge gap by the energy of the $L{=}N$ member of the CF-exciton mode~\cite{Jain97, Balram16d}.
 
We independently evaluate the Coulomb and $V_0{=}1$ pseudopotential gaps for many systems using the Monte Carlo method and extrapolate these gaps to the thermodynamic limit~\cite{SM}. By superposing these gaps, we extract the variational gap for an arbitrary combination of the Coulomb and $V_0$ pseudopotential. As shown in Fig.~\ref{fig: bosons_variational}, the neutral and charge gaps have different slopes as a function of $\delta V_0$, the former decaying faster. Thus, there is a region of parameter space  $0.43 {\leq} {-}\delta V_0{\leq}0.84$ where the neutral gap vanishes while the charge gap remains finite. This variational estimate of the FQHN phase boundaries is consistent with the ED results above.

\textbf{\textit{Symmetry breaking and geometric response.}} As an order parameter for the FQHN transition, we utilize the deformed Laughlin states which partly break rotational symmetry~\cite{Musaelian96, Qiu12}: 
\begin{equation}
\Psi_{\nu_{b}=1/2,\alpha}=\prod_{i<j}\left(z_i - z_j -\alpha\right)\left(z_i - z_j +\alpha\right),
\label{alpha}
\end{equation}
where we have suppressed the usual Gaussian factor and the parameter $\alpha$ controls the breaking of rotational symmetry by ``splitting'' the zeros of the wave function. For small $\alpha$, these wave functions, as well as related ones in Refs.~\cite{Haldane11, Qiu12}, describe an incompressible fluid rather than an FQHN. Nevertheless, we can use their variational energy to define a ``mean-field'' order parameter: we keep the interaction fully rotation-invariant, while we evaluate the expectation value of the energy for the anisotropic wave functions. For $\delta V_0 {>} \delta V_0^{\rm critical}$, we expect the energy minimum to correspond to $\alpha{=}0$, i.e., the isotropic Laughlin wave function, while for $\delta V_0 {<}\delta V_0^{\rm critical}$ the minimum should shift to a non-zero $\alpha^*{>}0$.  In the vicinity of a second-order phase transition, mean-field theory predicts a critical component $\beta{=}1/2$ for the order parameter $\alpha^* {\propto} (\delta V_0^{\rm critical} {-}\delta  V_0)^{\beta}$~\cite{You14}. 

\begin{figure}[bt]
    \centering    
    \includegraphics[width=0.98\linewidth]{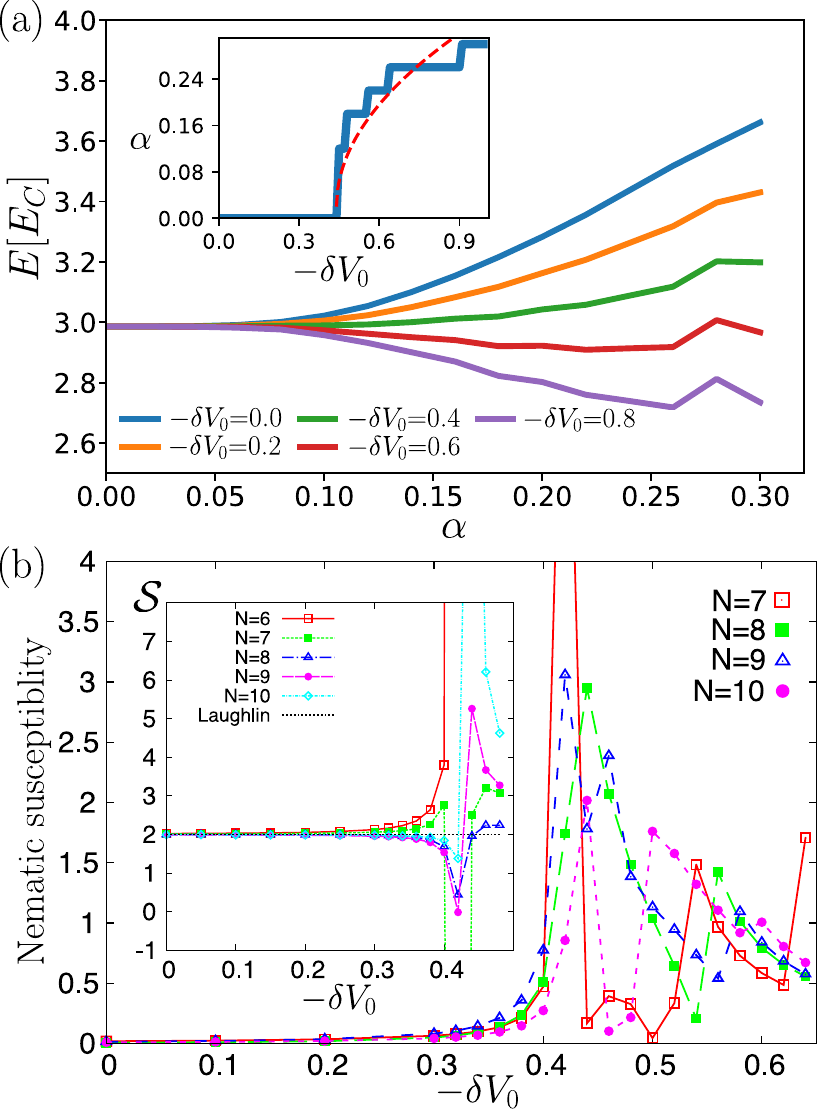}
    \caption{
    (a) Energy of the wave functions in Eq.~\eqref{alpha} for various $\delta V_0$. We discretize the $\alpha$ range in steps of $0.02\ell_B$. Inset shows the value of $\alpha$ which minimizes the total energy at each $\delta V_0$. The Coulomb energy and $V_0{=}1$ energy are calculated respectively for $N{=}15,20,25,30$ and the thermodynamic values are obtained from finite-size extrapolation. The red dashed line is the fitting of the data with the form of $\kappa\sqrt{\delta V_0^{\rm critical} {-} \delta V_0}$ where $\kappa{=}0.439$ and $\delta V_0^{\rm critical}{=}{-}0.44$. 
    (b) Nematic susceptibility from Ref.~\cite{Regnault17} for bosons at $\nu_b{=}1/2$ on the torus with a square unit cell for several system sizes indicated in the legend. As $V_0$ is reduced and the critical point is approached, we observe a sharp increase and strong finite-size fluctuations in the nematic susceptibility. Similar behavior is found for the shift $\mathcal{S}$ shown in the inset. We extract the shift from the Hall viscosity $\eta^A$ on the torus near the square aspect ratio. In the gapped FQH phase, the shift is quantized to the $\mathcal{S}{=}2$ value expected in the Laughlin state. Similar to the nematic susceptibility, as the FQHN critical point is approached, we observe large fluctuations in $\mathcal{S}$, which is no longer quantized.
    }
    \label{fig: BRS_suscep}
\end{figure}

This scenario is confirmed in Fig.~\ref{fig: BRS_suscep}(a), which shows the energy of $\Psi_{\nu_{b}{=}1/2,\alpha}$ states for the LLL-projected Coulomb interaction at several values of $\delta V_0$. The isotropic Laughlin state (recovered at $\alpha{=}0$) indeed yields the minimum of energy when $|\delta V_0|{\leq} 0.44{=}{-}\delta V_0^{\rm critical}$. This estimate of the FQHN transition point is in good agreement with our previous exact results for the closing of the neutral gap. Upon further reducing $V_0$, the lowest energy state appears at a nonzero $\alpha^*$ value. The scaling of $\alpha^*$ near criticality is presented in the inset of Fig.~\ref{fig: BRS_suscep}(a), which is consistent with the mean-field exponent 1/2. 

To characterize the geometric nature of the FQHN transition, we have computed the nematic susceptibility introduced in Ref.~\cite{Regnault17}, see Fig.~\ref{fig: BRS_suscep}(b). This calculation is performed in the torus geometry~\cite{Haldane85} by varying the aspect ratio. Fig.~\ref{fig: BRS_suscep}(a) shows that deep in the Laughlin phase the nematic susceptibility is zero while close to the transition point, $\delta V_{0}{=}{-}0.43$, it jumps to a non-zero value and thereafter varies erratically with system size. Moreover, we have computed the Hall viscosity $\eta^A$ on the torus~\cite{Avron95, Haldane09, Read11} as the system is tuned towards the FQHN critical point~\cite{SM}. For gapped FQH states, the Hall viscosity is quantized by the shift, $\eta^A{=}(\hbar \rho/4)\mathcal{S}$~\cite{Read09}, where $\rho$ is the fluid density. The extracted value of $\mathcal{S}$ is plotted in the inset of Fig.~\ref{fig: BRS_suscep}(b) as a function of $\delta V_0$. Deep in the gapped phase, $\delta V_0 {\gtrsim} {-}0.3$, the shift is robustly quantized to the Laughlin value $\mathcal{S}{=}2$. Near the transition but still in the gapped phase, the smaller systems depart from the quantized value, which is attributed to an increase in the correlation length. Finally, as we hit the FQHN critical point, the quantization of $\eta^A$ breaks down completely. In the SM \cite{SM}, we show that the onset of the transition is also signaled by the dynamical response of the system to a geometric quench.

\textbf{\textit{Experimental implications.}} The key ingredient of our model -- the softening of the $V_0$ pseudopotential of the Coulomb interaction -- can realistically arise due to the finite width of the 2DEG or screening by electrostatic gates. LL mixing, in particular, leads to a large reduction of $V_0$~\cite{Sodemann13, Peterson14, Simon13}. Moreover, the node in the single-particle wave functions in higher LLs tends to expose the magnetoroton mode in the long-wavelength limit~\cite{Jolicoeur17}, thus providing a natural setting for FQHNs~\cite{Fu21}. One of the main experimental challenges is distinguishing the FQHN from a CDW. Resonant inelastic Raman scattering~\cite{Pinczuk93, Du19,Liang2024} or surface acoustic waves~\cite{Kukushkin09} can map out the magnetoroton mode and confirm whether it closes in the long-wavelength limit. Recent advances in the scanning tunneling microscopy of FQH states~\cite{Xiaomeng2022, Coissard2022, Farahi2023, Hu23} could provide further insights into the FQHN formation, generalizing the previous observation of nematicity of free electrons in bismuth~\cite{Feldman16}.

Beyond solid-state systems, interactions in synthetic systems of alkaline atoms can be tuned by coupling to the highly excited Rydberg levels. The Rydberg blockade radius that simulates the $V_0$ interaction can be tuned in these platforms and both FQH and crystalline phases have been shown to arise in these models~\cite{Grusdt13, Grass18}. 
The long-range interactions in Rydberg-dressed atoms as well as dipolar gases~\cite{Defenu23} make them a promising platform for FQH physics~\cite{Burrello20}.  
Other platforms where bosonic FQH states can be stabilized are polaritons where an artificial magnetic field and LLs can be generated by rotating the medium through which light propagates~\cite{Otterbach10}. A suitably chosen medium could potentially allow to engineer a desired interaction~\cite{Knuppel2019, Bloch2022}. 

\textbf{\textit{Conclusions.}} In summary, we have presented a microscopic model that exhibits key features of the FQHN transition. Intriguingly, the model shows clearer FQHN signatures compared to short-range models, such as the one in Ref.~\cite{Regnault17}. We speculate that this is due to the tendency of the Coulomb interaction to ``pull down'' the magnetoroton mode below the spectral continuum in the long-wavelength limit, as seen in Fig.~\ref{fig: bosons_spectrum_gaps}. By contrast, in short-range models, the long-wavelength limit of the magnetoroton mode is clearly inside the continuum~\cite{Yang12b}. The necessary conditions for the gaplessness of the $L{=}2$ neutral excitation in short-range models have been derived in Ref.~\cite{Yang20Nematic}, and it would be interesting to generalize those results to long-range interactions. 

While we have provided multiple pieces of evidence for the FQHN critical point, the nature of the phase \textit{past} criticality is not fully understood. At large negative $\delta V_0$, we expect a CDW to become the ground state. On the sphere, the study of CDW phases is hindered by frustration effects, requiring very large systems to observe the expected closing of the charge gap. On a torus, by varying the aspect ratio, we have indeed found a proximate phase with a manifold of ground states consistent with CDW ordering~\cite{SM}. Thus, the FQHN phase found above could be proximate to a CDW.

One important question is: How general are the results above? In the SM~\cite{SM}, we show that fermions at $\nu{=}1/3$ behave similarly to bosons at $\nu_b{=}1/2$ considered above. Moreover, other incompressible states, in particular $\nu{=}2/5$ and $3/7$ Jain states, also have collective modes ``exposed'' below the continuum of the spectrum~\cite{Balram13, Jain14}. We leave their study to future work. We note, however, that the FQHN presented above could potentially be realized even beyond non-interacting CF states. For example, we have checked that bosons at $\nu_b{=}1$, which realize the Moore-Read state~\cite{Moore91, Cooper01, Regnault03, Regnault07, Sharma23}, support a similar phenomenology.  Beyond the nematic instability of incompressible FQH states, it would be interesting to explore a \emph{compressible} nematic which could arise in third and higher Landau levels \cite{Lilly99, Lilly99b, Fradkin00,ye2024}, e.g., due to the quantum (or thermal) melting of a stripe state \cite{Fradkin99, Qian17, Fu21} or a Pomeranchuk instability of the compressible $\nu{=}1/2$ state \cite{Lee18, You16}. 

\textbf{\textit{Acknowledgments}.} We thank Bo Yang, Joseph Maciejko, Mikael Fremling, Nicolas Regnault, and, in particular, Zhao Liu for useful discussions. S.P., J.T., and Z.P. acknowledge support from the Leverhulme Trust Research Leadership Award RL-2019-015. A.C.B. and Z.P. thank the Royal Society International Exchanges Award IES$\backslash$R2$\backslash$202052 for funding support. This research was supported in part by the International Centre for Theoretical Sciences (ICTS) for participating in the program - Condensed Matter meets Quantum Information (code: ICTS/COMQUI2023/9). This research was supported in part by grants NSF PHY-1748958 and PHY-2309135 to the Kavli Institute for Theoretical Physics (KITP), and DMR-2225920 at the University of Illinois (EF). This work was undertaken on ARC3 and ARC4, part of the High-Performance Computing facilities at the University of Leeds, UK, and on the Nandadevi supercomputer, which is maintained and supported by the Institute of Mathematical Science's High-Performance Computing Center, India. Some of the numerical calculations were performed using the DiagHam libraries~\cite{DiagHam}. ACB thanks the Science and Engineering Research Board (SERB) of the Department of Science and Technology (DST) for funding support via the Mathematical Research Impact Centric Support (MATRICS) Grant No. MTR/2023/000002.

\newpage 
\cleardoublepage

\onecolumngrid
\begin{center}
\textbf{\large Supplemental Online Material for ``Microscopic Model for Fractional Quantum Hall Nematics'' }\\[5pt]

\begin{center}
 {\small Songyang Pu$^{1,2}$, Ajit C. Balram$^{3,4}$, Joseph Taylor$^{1}$, Eduardo Fradkin$^{5,6}$ and Zlatko Papi\'c$^{1}$ }  
\end{center}

\begin{center}
{\sl \footnotesize
$^{1}$School of Physics and Astronomy, University of Leeds, Leeds LS2 9JT, United Kingdom

$^{2}$Department of Physics and Astronomy, The University of Tennessee, Knoxville, TN 37996, USA

$^{3}$Institute of Mathematical Sciences, CIT Campus, Chennai 600113, India

$^{4}$Homi Bhabha National Institute, Training School Complex, Anushaktinagar, Mumbai 400094, India
    
$^{5}$Department of Physics, University of Illinois at Urbana-Champaign,
1110 West Green Street, Urbana, Illinois 61801, USA

$^{6}$Anthony J. Leggett Institute for Condensed Matter Theory,
University of Illinois at Urbana-Champaign, 1110 West Green Street, Urbana, Illinois 61801, USA
}
\end{center}

\begin{quote}
{\small In this Supplemental Material, we present further details on the methodology used in the main text and additional results that support our conclusions: (i) we discuss in detail the different types of gaps that are studied in the main text; (ii) we introduce three classes of variational wave functions that we have used to demonstrate the existence of FQH nematic transition in system sizes beyond exact diagonalization; (iii) we show that the FQH nematic transition can be probed by studying the dynamical response of the system to the geometric quench; (iv) we show our main result about the existence of FQHN critical point is insensitive to the boundary condition by calculating the spectrum in disk geometry; (v) we outline our procedure for calculating the Hall viscosity, which simplifies the method in Ref.~\cite{Read11};  (vi) we provide a detailed study of the charge-density-wave (CDW) phase in the torus geometry; (vii) we demonstrate that our conclusions in the main text also hold for fermions at filling factor $\nu{=}1/3$. 
}
\end{quote}
\end{center}

\vspace*{0.4cm}

\twocolumngrid 

\setcounter{equation}{0}
\setcounter{figure}{0}
\setcounter{table}{0}
\setcounter{page}{1}
\setcounter{section}{0}
\makeatletter
\renewcommand{\theequation}{S\arabic{equation}}
\renewcommand{\thefigure}{S\arabic{figure}}
\renewcommand{\thesection}{S\Roman{section}}
\renewcommand{\thepage}{\arabic{page}}
\renewcommand{\thetable}{S\arabic{table}}

\section{Definitions of various gaps}\label{app:gaps}

In the main text, we have computed three types of gaps to characterize the fractional quantum Hall nematic (FQHN) transition: (i) the neutral gap is defined as the difference in the lowest two energies at the ground state flux, i.e., $2Q{=}2N{-}2$ for the $\nu_b{=}1/2$ Laughlin state; (ii) the charge or transport gap, i.e., the energy to create an individual quasihole (occurs at flux $2Q{+}1$) and quasiparticle (occurs at flux $2Q{-}1$) on top of the ground state; (iii) 
the gap $\Delta^{L{=}N}$ -- the energy difference between the ground state and the lowest-lying state with orbital angular momentum $L{=}N$ -- at the ground state flux. These three types of gaps were computed for the accessible system sizes $N{=}6{-}12$ bosons 
and their extrapolation to the thermodynamic limit in $1/N$ was performed. 
For the charge gap, it is important to include the neutralizing background charge correction~\cite{Morf02, Jain97} Since we are considering modified Coulomb interaction, the total electrostatic energy is evaluated according to the general formula in the Supplementary Material of Ref.~\cite{Balram20b}, which we reproduce here for convenience:
\begin{eqnarray}
    E_\mathrm{corr}(2Q) = -\frac{1}{2}\mathcal{C}_{2Q} \left(N^2 - n_q^2 e_*^{2}\right), \\ \mathcal{C}_{2Q} = \sum_{m=0}^{2Q} V_m \frac{4Q-2m+1}{(2Q+1)^2},
\end{eqnarray}
where $n_q$ is an integer taking values 0 (for the ground state) or ${\pm} 1$ for the quasiparticle and quasihole excitations that carry charge $e_*{=}e/2$ in our case. The energy correction explicitly depends on the number of flux quanta $2Q$ (we have assumed we are in the lowest Landau level). The average charging energy of the full LL per pair, $\mathcal{C}$, is determined by all pseudopotentials $V_m$, which includes any modification to the $V_0$. Furthermore, to minimize the finite-size effects, we multiply all energies with the density-correction factor $\sqrt{2Q\nu_{b}/N}$~\cite{Morf86}. In the case of charge gap, we do not density-correct the individual quasihole and quasiparticle energies but only apply the rescaling at the end. 

We use the gap $\Delta^{L{=}N}$ as a cross-check for the extrapolated value of the charge gap in the thermodynamic limit. The $L{=}N$ magnetoroton state is formed by a quasiparticle and a quasihole on opposite poles of the sphere. As the charge gap is expected to remain finite across the FQHN transition, the quasiparticle and quasihole are localized objects with a few $\ell_B$ in diameter. By contrast, their interaction scales inversely with the diameter of the sphere, $2R{\propto}\sqrt{N}\ell_B$. Thus, in the thermodynamic limit, the quasiparticle and quasihole are infinitely far from each other and thus the interaction between them vanishes in this limit and they can be regarded as independent entities, the sum of whose energies is equivalent to the charge gap.

\section{Variational wave functions for the magnetoroton mode}
\label{app: variational}

For completeness, we provide an overview of three construction schemes of variational wave functions that we have used to approximate the neutral collective mode in the main text. 

\subsection{Jack polynomial wave functions and Girvin-MacDonald-Platzman ansatz}\label{sec:jack}

Accurate model wave functions for the magnetoroton mode of the Laughlin state can be constructed in the framework of Jack polynomials~\cite{Bernevig08}. The construction applies equally to bosons or fermions, and below we briefly review the fermionic version, commenting on how to adapt it to bosons at the end.  The Laughlin wave function at $\nu{=}1/3$ is a single Jack polynomial characterized by the so-called root partition $\lambda{\equiv} \{100100100100{\cdots} 1001\}$. The root is a particular Fock basis state from which all the other Fock states that carry non-zero weight in the Laughlin wave function can be obtained via ''squeezing''~\cite{Bernevig08a}: a two-particle process where one particle moves from orbital $m_1$ to $m_1^{\prime}$ and another from $m_2$ to $m_2^{\prime}$, where $m_1{<}m_1^{\prime} {\leq} m_2^{\prime}{<}m_2$, and $m_1{+}m_2{=}m_1^{\prime}{+}m_2^{\prime}$. 

The root partition encodes the following clustering property, unique to the Laughlin $1/3$ state: no more than a single electron can exist in three consecutive orbitals. The collective mode in the limit of momentum $k{\to} 0$ consists of two quasielectron-quasihole pairs, forming a quadrupole. As the momentum increases, a dipole moment develops with the separation of one quasielectron-quasihole pair. This idea can be translated into an explicit set of root partitions for different states comprising the magnetoroton mode: 
\begin{eqnarray}\label{eq: laughlinwfs}
\nonumber L=2: \quad && \underset{\bullet}{1} \underset{\bullet}{1} 0 \underset{\circ}{0} \underset{\circ}{0} 0 1 0 0 1 0 0 1 0 0 1 0 0 1 \cdots 0 0 1 0 0 1 \\
\nonumber L=3: \quad && \underset{\bullet}{1} \underset{\bullet}{1} 0 \underset{\circ}{0} 0 1 0 \underset{\circ}{0} 0 1 0 0 1 0 0 1 0 0 1 \cdots 0 0 1 0 0 1 \\
&& \vdots \\
\nonumber L=N: \quad && \underset{\bullet}{1} \underset{\bullet}{1} 0 \underset{\circ}{0} 0 1 0  0 1 0 0 1 0 0 1 0 0 1 \cdots 0 0 1 0 0 1 \underset{\circ}{0}, 
\end{eqnarray} 
where the filled circle indicates the position of a quasielectron, while the empty circle denotes the quasihole.  The configurations above are labeled by their total angular momentum $L$ on the sphere, and the smallest momentum that can be created is $L{=}2$. 

The wave functions for the magnetoroton states $|\psi^L_\lambda\rangle$ are ultimately determined by imposing the following conditions~\cite{Yang12b}
\begin{eqnarray}\label{eq: hwt}
|\psi^L_\lambda\rangle=\sum_{\mu\preceq\lambda}a_\mu \mathrm{sl}_\mu, \\
L^+|\psi^L_\lambda\rangle=0,\\
\hat{V}_1 \hat c_{Q} \hat c_{Q-1}|\psi_\lambda^L\rangle=0.
\end{eqnarray}
Here, $\mathrm{sl}_\mu$ denotes a Slater determinant with an orbital occupation pattern given by the partition $\mu$. The first condition thus means that we are looking for states that are obtained by squeezing from one of the roots $\lambda$ in Eq.~\eqref{eq: laughlinwfs}, which introduces a partial ordering among the partitions, denoted by $\preceq$. The second condition states that we are only interested in rotationally invariant states.  These two conditions uniquely fix the coefficients $a_\mu$ of the Laughlin ground state wave function. For the excited magnetoroton states, to find $a_\mu$ we require the third, energetic condition, involving the $\hat V_1$ pseudopotential operator and electron annihilation operators $\hat c_m$, which singles out the state with quasiparticles piled up at the north pole of the sphere. The resulting states are confirmed to be excellent approximations to the eigenstates of Coulomb interaction~\cite{Yang12b}. Moreover, the construction can be readily generalized to other types of states that have a Jack polynomial description and a parent Hamiltonian, e.g., the Read-Rezayi sequence~\cite{Read99}.

The conditions above can be straightforwardly modified to yield the magnetoroton states of the bosonic Laughlin state at $\nu_b{=}1/2$. The clustering condition now stipulates that there should be no more than one boson in every two orbitals. The first condition in Eq.~\eqref{eq: hwt} should then be viewed as an expansion over permanents (rather than Slater determinants), while the last condition needs to be replaced with $\hat V_0$ pseudopotential. Finally, the root configurations need to be divided with a Jastrow factor, resulting in bosonic roots such as $2000101010{\cdots} 10101$ (for $L{=}2$), $2001001010{\cdots} 10101$ (for $L{=}3$), etc. With these modifications, the construction of bosonic magnetoroton states is fully analogous to the fermionic case.

\begin{figure}[bt]
    \centering    
    \includegraphics[width=0.95\linewidth]{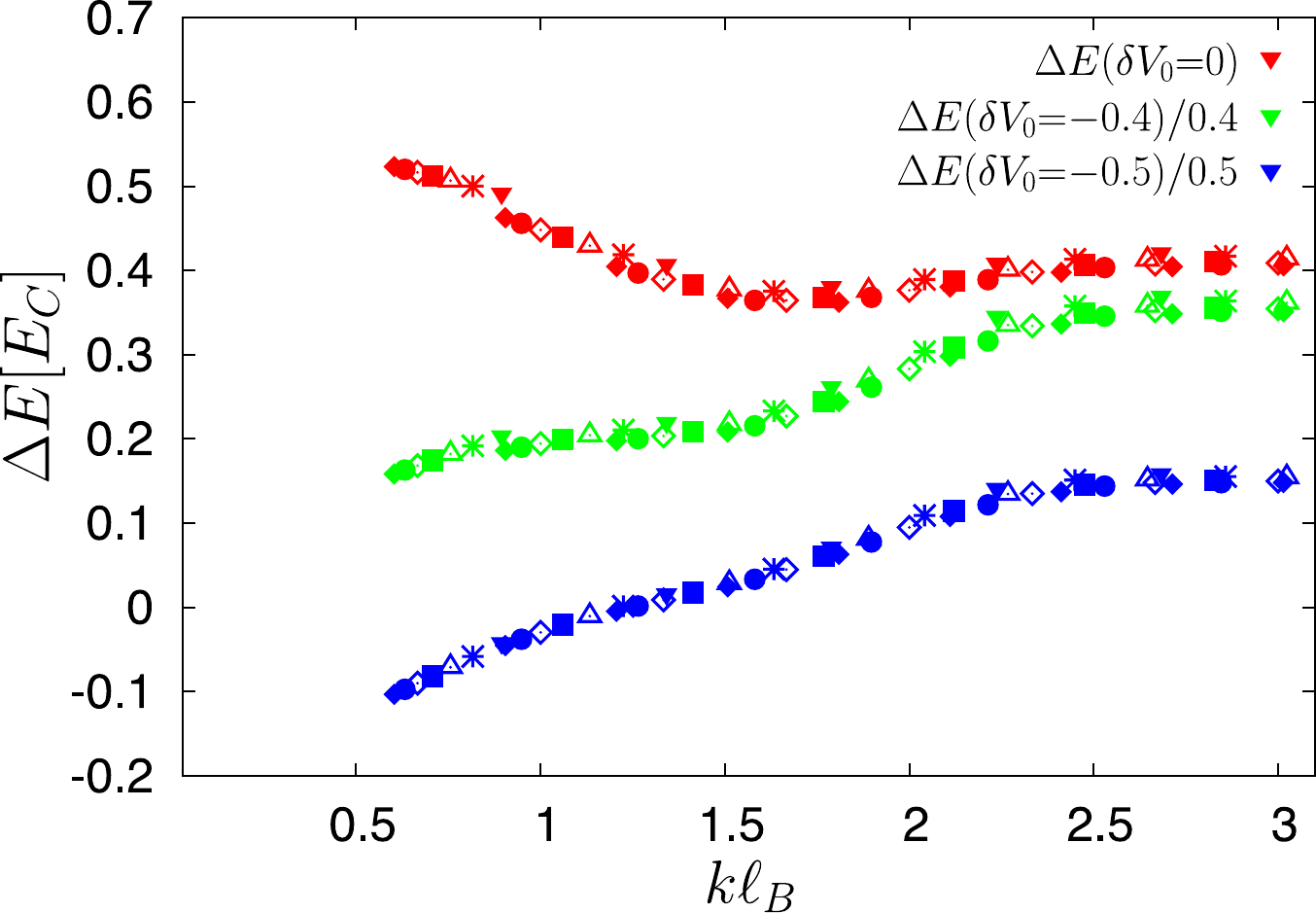}
    \caption{
     Variational energies for the Jack polynomial approximations to the collective mode states from Ref.~\cite{Yang12b} for the bosonic $\nu_{b}{=}1/2$ state. For $\delta V_0 {\neq} 0$ cases, we have rescaled the energies with  $|\delta V_0|$ for clarity. 
    }
    \label{fig: bosons_variational_jack}
\end{figure}

We evaluate the variational energy of the Jack wave functions in Fig.~\ref{fig: bosons_variational_jack} for three values of $\delta V_0$. This shows, within the variational approximation, that the collective mode becomes gapless in the $k{\to} 0$ limit, supporting the results of exact numerics in the main text.

It turns out that the elaborate construction presented above, in special cases $L{=}2,3$, reduces to the much simpler single-mode approximation (SMA) ansatz written down by Girvin, MacDonald, and Platzman~\cite{Girvin85}. On the sphere, the SMA ansatz for a state with total angular momentum $L$ is obtained by acting on the ground state with the projected density operator $\hat{\rho}_{L, M}$~\cite{He94} on the ground state
\begin{eqnarray}\label{eq: smawf}
|\psi^{\rm SMA}_{LM}\rangle {=} \hat{\rho}_{L,M} |\psi_0\rangle,
\end{eqnarray}
where $\hat{\rho}_{L, M} {=} \sum_m C^{Q, L, Q}_{m{+}M, M,m} c_{m{+}M}^\dagger c_m$, with $C^{Q, L, Q}_{m{+}M, M,m}$ denoting the Clebsch-Gordon coefficient and, due to the rotational invariance, we can set $M{=}L$. 

At $L{=}1$, the density operator in Eq.~(\ref{eq: smawf}) reduces to the angular momentum raising operator~\cite{He94}, hence it annihilates the ground state, which is a singlet ($L{=}0$). Thus, the $L{=}1$ SMA trial state identically vanishes, which is also the case for the Jack construction. Moreover, at $L{=}2,3$, the SMA wave functions only involve the elements of the basis squeezed from the same root partition that defines the Jack model wave functions in Eq.~\eqref{eq: laughlinwfs}. Hence, for the Laughlin $1/3$ state, the equivalence between the two sets of wave functions can be proven for $L{=}2,3$~\cite{Yang12b}. Moreover, both of these states are also equivalent to the CF exciton states discussed 
below. For $L{>}3$ the SMA wave functions contain unsqueezed basis components with respect to the root partitions used in the Jack model wave functions, thus the final states are no longer equivalent. The accuracy of the SMA ansatz in Eq.~\eqref{eq: smawf}, however, is numerically found to deteriorate rapidly at larger values of $L$~\cite{Repellin14}, unlike the Jack wave functions which remain accurate over the entire range $2{\leq} L {\leq} N$. 

\subsection{LLL-projected GMP wave function}\label{app:gmp}

The original SMA ansatz, Eq.~(\ref{eq: smawf}), is not convenient for evaluation in large systems. In this subsection, we present an equivalent real-space form of the GMP ansatz that we have used to calculate the $L{=}2$ gap. We consider general $\nu{=}1/m$ Laughlin states, which can be either bosonic for even $m$ or fermionic for odd $m$.

In first quantization, the GMP wave function [Eq.~\eqref{eq: smawf}] on the sphere takes the form~\cite{Girvin85}
\begin{align}
    \Psi_{\rm GMP}(L)
    &=\sum_{i=1}^N\mathcal{P}_{\rm LLL}Y_{L,M}(u_i,v_i)\prod_{j<k}(u_jv_k-u_kv_j)^m,
\end{align}
where $Y_{L,M}(u_i,v_i)$ are the spherical harmonics. Due to rotational invariance, we can set $M{=}{-}L$ for simplicity, thereby $Y_{L,{-}L}(u_i,v_i)\propto v_i^L\bar{u}_i^L$. To implement the LLL projection exactly and efficiently, we apply the technique developed in Ref.~\cite{Pu23} to obtain
\begin{align}
\nonumber \Psi_{\rm GMP}(L)=&\mathcal{P}_{\rm LLL}\sum_i Y_{L,-L}\left(u_i,v_i\right)\prod_{j<k}\left(u_jv_k-u_kv_j\right)^m\\ \nonumber
=&\mathcal{P}_{\rm LLL}\sum_i v_i^L\bar{u}_i^{L}\prod_{j<k}\left(u_jv_k-u_kv_j\right)^m\\ 
\nonumber =&\sum_i v_i^L\left(\frac{\partial}{\partial u_i}\right)^{L} \prod_{j<k}\left(u_jv_k-u_kv_j\right)^m\\
=&\sum_i v_i^L\prod_{j<k}\left(u_jv_k-u_kv_j\right)^mP_i(L,m).
\end{align}
In the last line, we defined a new function $P_i(L,m)$
\begin{eqnarray}
\notag  P_i(L,m) &\equiv& \prod_{j<k}\left(u_jv_k-u_kv_j\right)^{-m} \\ &\times& \left(\frac{\partial}{\partial u_i}\right)^{L}\prod_{j<k}\left(u_jv_k-u_kv_j\right)^m.    
\end{eqnarray}
This newly defined function can be evaluated through the recursion relations
\begin{equation}
    P_i(L,m)=
    \begin{cases}
    1, & L=0\\
    \left(mh_i(1)+\frac{\partial}{\partial u_i}\right)P_i(L-1,m), & L\geq 1
    \end{cases}.
\end{equation}
Here, the function $h_i(L)$ is defined as
\begin{equation}
    h_i(L)=\sum_{j\neq i}\left(\frac{v_j}{ u_iv_j-u_jv_i}\right)^L.
\end{equation}
It has the following property under the action of a derivative
\begin{equation}
    \frac{\partial}{\partial u_i}h_i(L)=-Lh_i(L+1).
\end{equation}

For the bosonic Laughlin state with $m{=}2$, we calculate the energy for the $V_0$ interaction, which requires setting $\vec{\Omega}_1 {=} \vec{\Omega}_2$ in the wave function. In this special case, the GMP wave function can be written exactly as
\begin{eqnarray}
   \nonumber  \Psi_{\rm GMP}^{\vec{\Omega}_1 = \vec{\Omega}_2}(L) \hspace{-0.1cm}&=& \hspace{-0.1cm} \sum_{i=1,2}2v_i^Lv_{3-i}^2 \\
   && \hspace{-0.5cm} \prod_{j<k,j+k>3}\left(u_jv_k-u_kv_j\right)^m\tilde{P}_i(L-2,2),  \quad \quad \quad 
\end{eqnarray}
in which $\tilde{P}$ is computed through
\begin{equation}
    \tilde{P}_i(L,2)=
    \begin{cases}
    1, & L=0\\
    \left(2\tilde{h}_i(1)+\frac{\partial}{\partial u_i}\right)\tilde{P}_i(L-1,2), & L\geq 1.
    \end{cases}
\end{equation}
Here, the function $\tilde{h}_i(L)$, for $i{=}1,2$, is defined as
\begin{equation}
    \tilde{h}_i(L)=\sum_{j\geq 3}\left(\frac{v_j}{ u_iv_j-u_jv_i}\right)^L.
\end{equation}

\subsection{Composite fermion exciton wave function}\label{sec:cfexc}

The bosonic Laughlin state occurs when the effective flux seen by electrons is $2Q^{*}{=}N{-}1$. We will keep using the language of composite fermions even though the composite object of an electron and a single vortex here is a boson. These states can be viewed as (CF state) ${\times}$ ($\nu{=}1$ integer quantum Hall state).

The exact form of the $\nu_{b}{=}1/2$ composite fermion exciton wave function for total orbital angular momentum $L$ and $L_{z}{=}M$ is given as
\begin{widetext}
\begin{eqnarray}
   \Psi^{\rm CFE}_{1/2}(L,M; \{ \vec{\Omega_{i}} \} ) \hspace{-0.1cm} &{=}& \hspace{-0.1cm} \prod_{i\neq j} (u_{i} v_{j} {-} u_{j} v_{i})\sum_{k=1}^N (-1)^k \sum_{m_{1}} \mathcal{C}^{L,M}_{Q^{*},m_{1};Q^{*}{+}1,M{-}m_{1}} \Tilde{\Psi}_{m_{1},k}(\{ \vec{\Omega_{i}} \}),  
\end{eqnarray}
where $\mathcal{C}^{L, M}_{Q^{*},m_{1}; Q^{*}{+}1, M{-}m_{1}}$ is the Clebsch-Gordan coefficient that comes about from adding the angular momentum of the constituent CF hole (${-}m_{1}$) and CF particle ($m_{2}{=}M{-}m_{1}$). The wave function $\Tilde{\Psi}$ is given as
\begin{equation}
    \Tilde{\Psi}_{m_{1},k}(\vec{\Omega_{1}},\vec{\Omega_{2}},\vec{\Omega_{3}},{\dots},\vec{\Omega_{N}})=\Tilde{Y}_{Q^{*}, M{-}m_{1}} (\vec{\Omega_{k}}) \Phi^{\rm hole}_{m_{1}}(\vec{\Omega_{1}},\vec{\Omega_{2}},{\dots},\vec{\Omega_{k-1}},\vec{\Omega_{k+1}},\vec{\Omega_{N}}),
\end{equation}
where $\Phi^{\rm hole}_{m_{1}}$ is the $(N{-}1){\times}(N{-}1)$ Slater determinant state that describes a hole at $\nu{=}1$ which is obtained by leaving the ${-}m_{1}$ orbital unoccupied and $\Tilde{Y}_{q, m} (\vec{\Omega_{k}})$ is given as
\begin{equation}
    \Tilde{Y}_{q, m} (\vec{\Omega_{k}}) = \mathcal{N}_{q, m} ({-}1)^{q{-}m}  u^{q{+}m}_{k}v^{q{-}m}_{k} \left[ \sum_{j\neq k}{1\over u_kv_j-v_ku_j} \left[\binom{2q{+}1}{q+m}u_j v_k + \binom{2q{+}1}{q+m+1} u_k v_j\right] \right],
\end{equation}
where $\mathcal{N}_{q, m}$ is a normalization constant given by
\begin{equation}
    \mathcal{N}_{q, m} = \sqrt{\frac{2q{+}3}{4\pi}\frac{(q{+}1{-}m)!(q{+}1{+}m)!}{(2q{+}1)!}}.
\end{equation}

\subsubsection{CF exciton wave function for $\vec{\Omega_{1}}{=}\vec{\Omega_{2}}$}

The exact form of the $\nu_{b}{=}1/2$ CF exciton wave function for total orbital angular momentum $L$ and $L_{z}{=}M$ when coordinates of two of the bosons are identically set equal to each other, which is required to evaluate its energy for the $\delta$-function interaction~\cite{Chang05b}, is given as
\begin{equation}
    \Psi^{\rm CFE}_{1/2}(L,M;\vec{\Omega_{2}},\vec{\Omega_{2}},\vec{\Omega_{3}},{\dots},\vec{\Omega_{N}})  {=} \prod_{k{=}3}^{N} (u_{2} v_{k} {-} u_{k} v_{2}) \prod_{2{\leq}j{<}k{\leq}N} (u_{j} v_{k} {-} u_{k} v_{j}) \sum_{m_{1}} \mathcal{C}^{L,M}_{Q^{*},m_{1}{-}M;Q^{*}{+}1,m_{1}} \Tilde{\Psi}_{m_{1}}(\vec{\Omega_{2}},\vec{\Omega_{2}},\vec{\Omega_{3}},{\dots},\vec{\Omega_{N}}),
\end{equation}
where $\mathcal{C}^{L, M}_{Q^{*},m_{1}{-}M; Q^{*}{+}1,m_{1}}$ is the Clebsch-Gordan coefficient that comes about from adding the angular momentum of the constituent CF particle ($m_{1}$) and CF hole (${-}m_{2}{=}M{-}m_{1}$) [Note, the CF hole has angular momentum of opposite sign so that $m_{1}{-}m_{2}{=}M$.]. The wave function $\Tilde{\Psi}$ is given as
\begin{equation}
    \Tilde{\Psi}_{m_{1}}(\vec{\Omega_{2}},\vec{\Omega_{2}},\vec{\Omega_{3}},{\dots},\vec{\Omega_{N}})=2\Tilde{Y}_{Q^{*}, m_{1}} (\vec{\Omega_{2}}) \Phi^{\rm hole}_{m_{1}}(\vec{\Omega_{2}},\vec{\Omega_{3}},{\dots},\vec{\Omega_{N}}) ,
\end{equation}
where $\Phi^{\rm hole}_{m_{1}}$ is the $(N{-}1){\times}(N{-}1)$ Slater determinant state that describes a hole at $\nu{=}1$ which is obtained by leaving the $M{-}m_{1}$ orbital unoccupied and $\Tilde{Y}_{q, m} (\vec{\Omega})$ is given as
\begin{equation}
    \Tilde{Y}_{q, m} (\vec{\Omega}) =  \sqrt{\frac{2q{+}3}{4\pi}\frac{(q{+}1{-}m)!(q{+}1{+}m)!}{(2q{+}1)!}} ({-}1)^{q{+}1{-}m}  u^{q{+}m{+}1} v^{q{-}m{+}1}  \binom{2q{+}2}{q{+}1{-}m} .
\end{equation}
\end{widetext}

\section{Geometric quench dynamics}\label{app:dyn}

\begin{figure}[bt]
    \centering
    \includegraphics[width=\linewidth]{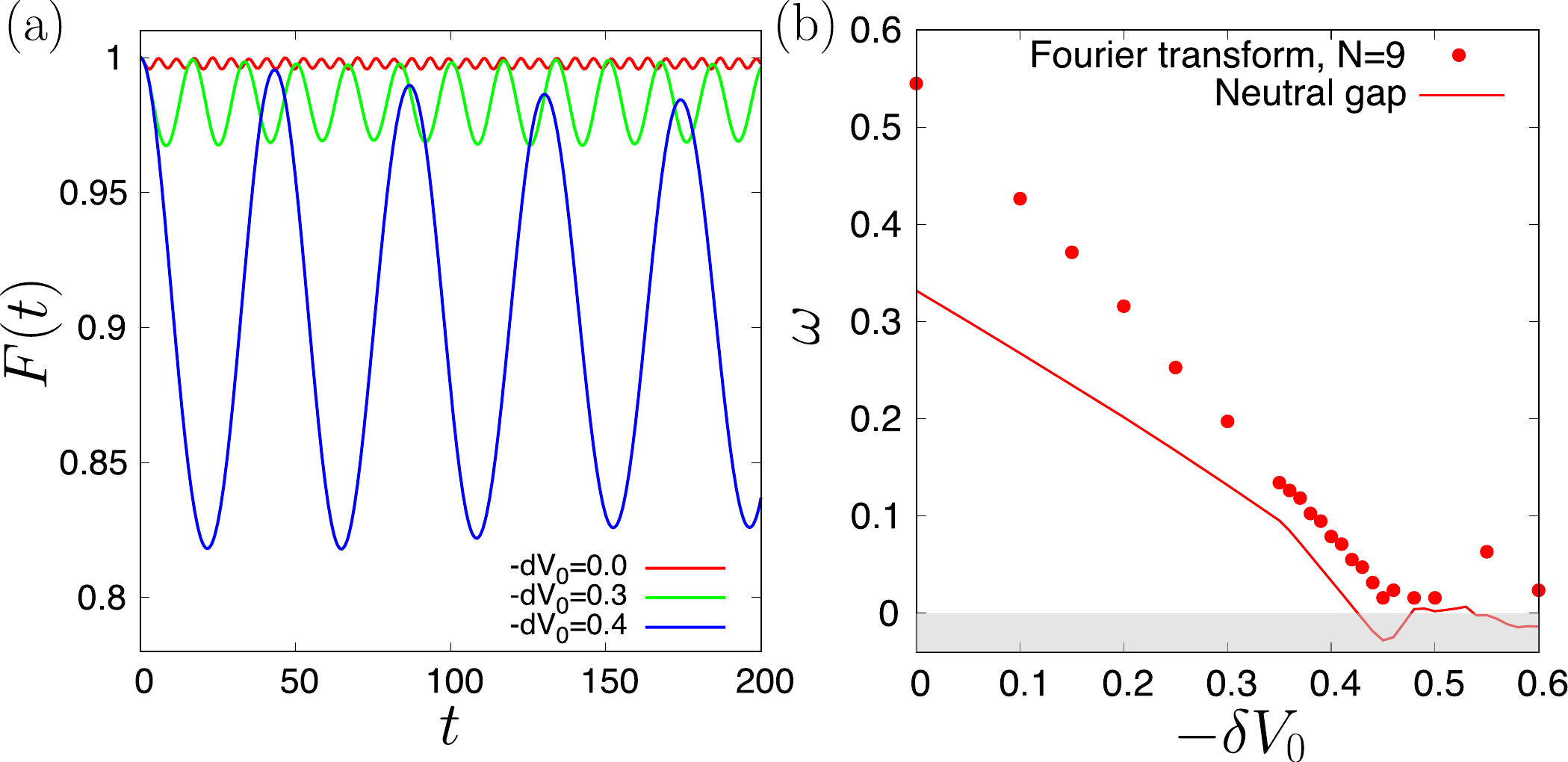}
    \caption{Geometric quench dynamics for bosons at $\nu_b{=}1/2$ in the vicinity of the FQHN critical point. (a) The fidelity for the FQH ground state for three values of the added $\delta V_0$ pseudopotential (indicated in the legend). All data is for a fixed system size $N{=}9$ and weak anisotropy $V_{0,2}{=}0.01$. (b) The frequency of the oscillations in panel (a), extracted via Fourier transform and plotted as a function of $\delta V_0$. For comparison, the line shows the neutral gap in the thermodynamic limit from the main text. 
    }
    \label{fig:dyn}
\end{figure}

The spectral properties discussed in the main text can be conveniently probed by quenching the FQH system out of equilibrium. We consider a particular type of ``geometric'' quench where the isotropic FQH state evolves under unitary dynamics generated by a weakly \textit{anisotropic} Hamiltonian~\cite{Liu18, Liu20}. Anisotropy induces quadrupolar deformations of the FQH fluid, which breaks the full rotational invariance on the sphere and couples the FQH ground state with the (isotropic) $L{=}2$ state of the GMP mode. In the spherical geometry, quadrupolar anisotropy for bosons can be conveniently modeled by adding a small amount of the anisotropic pseudopotential $V_{0,2}$, introduced in Ref.~\cite{Yang16}.  In Fig.~\ref{fig:dyn}(a) we present the geometric quench dynamics for bosons at $\nu_b{=}1/2$ for three values of the (isotropic) short-range pseudopotential $\delta V_0{=}0$, ${-}0.3$ and ${-}0.4$, approaching the FQHN transition. For all three points, we induce the dynamics by adding a fixed amount of $V_{0,2} {=} 0.01$. The dynamics is characterized by the fidelity, $F(t) {=} |\langle \psi(t) | \psi(0)\rangle|^2$, where $\psi(0)$ is the initial (isotropic) FQH ground state and $|\psi(t)\rangle{=}\exp({-}i t H^\prime) |\psi(0)\rangle$ is the state after time $t$ (we set $\hbar{=}1$). The prime on the Hamiltonian indicates that the interaction includes not only the modified $V_0$ pseudopotential but also the anisotropic component $V_{0,2}$. In all three cases, the dynamics are approximately harmonic and can be viewed as a two-level system. At the FQHN critical point itself, the harmonic dynamics are a finite-size effect that occurs due to the discreteness of the spectrum at fixed system size $N$. 

The fidelity in Fig.~\ref{fig:dyn}(a) oscillates with an increasing period as the FQHN point is approached. To quantify this behavior, in Fig.~\ref{fig:dyn}(b) we perform a Fourier transform on this data and plot the extracted frequency as a function of $\delta V_0$. For reference, we also include the extrapolated neutral gap at $N{\to}\infty$ from the main text. 
At small values of $\delta V_0$, i.e., deep in the Laughlin phase, the oscillation frequency is significantly higher than the neutral gap. This is because the neutral gap is determined by the lowest energy excitation, while the $L{=}2$ GMP state is higher in energy at the pure Coulomb point. As $V_0$ is reduced, the $L{=}2$ state becomes the lowest excited state around $\delta V_0 {\approx} {-}0.35$, corresponding to the kink in the neutral gap curve. For $\delta V_0{\gtrsim} {-}0.35$, the oscillation frequency is in good agreement with the neutral gap, with the small discrepancy between the two attributed to finite-size extrapolation. 

\section{Disk geometry}\label{app:disk}

In the main text, we have focused on closed geometries, sphere
 and torus, which are most convenient for studying the system's bulk properties and their extrapolation to the thermodynamic limit. In particular, we showed that our main conclusions about the existence of the FQHN critical point are largely insensitive to the choice of boundary condition. Hence, for sufficiently large systems, we expect these conclusions to hold also on a finite disk, which describes more closely the real physical system that necessarily has a boundary.  

\begin{figure}[tb]
    \centering
    \includegraphics[width=0.9\linewidth]{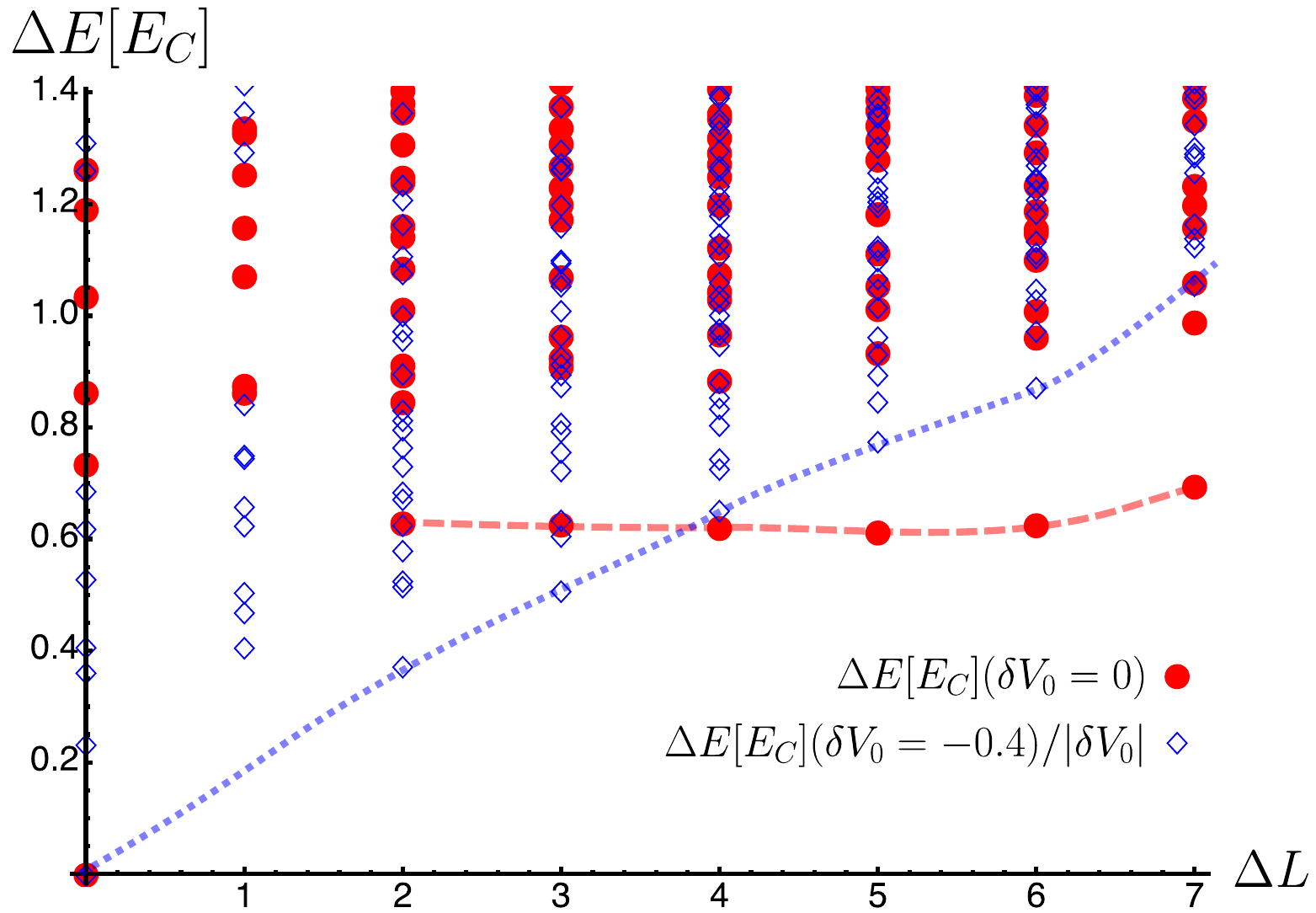}
    \caption{Energy spectrum for bosons at $\nu_b{=}1/2$ on the disk for pure Coulomb interaction (red circles) and softened Coulomb interaction with $\delta V_0{=}{-}0.4$ (blue diamonds). Data is for $N{=}7$ bosons in 13 orbitals with the center-of-mass momentum equal to zero. The spectrum is plotted relative to the momentum of the Laughlin ground state (see text).
    Similar to the sphere and torus, the softening of $V_0$ pseudopotential makes the magnetoroton mode (dashed line) go soft in the long-wavelength limit (dotted line). The lines are a guide to the eye. }
    \label{fig:disk}
\end{figure}

In Fig.~\ref{fig:disk} we confirm that the FQHN critical point is also realized on a finite disk in a similar parameter regime $\delta V_0 {=}{-}0.4$. The FQH Hamiltonian on a disk conserves the total angular momentum $L_z$ about the perpendicular $z$-axis and the bosonic Laughlin state at $\nu_b{=}1/2$ occurs in the sector $L_z^0{=}N(N{-}1)$. However, block-diagonalization of the Hamiltonian in $L_z$-invariant sectors is insufficient to reveal the neutral spectrum~\cite{Yang2019Disk}; instead, it is necessary to consider the approximately conserved center-of-mass angular momentum $\hat M_\mathrm{COM}$:
\begin{eqnarray}
    \hat M_\mathrm{COM} = \hat B^\dagger \hat B, \quad \hat B = \frac{1}{N}\sum_i \hat R_i,
\end{eqnarray}
where $\hat R_i$ is the guiding center coordinate of the $i$th particle. This operator can be expressed in second quantization with the help of Clebsch-Gordan coefficients on the disk, which can be found, e.g., in Ref.~\cite{Kang17}. For sufficiently small values of $L_z$, the spectrum of $\hat M_\mathrm{COM}$ is quantized in terms of integers~\cite{Yang2019Disk}. The collective mode of a FQH state can be resolved by projecting the Hamiltonian into a subspace of states with eigenvalue 0 of $\hat M_\mathrm{COM}$ and then plotting the energies as a function of $\Delta L {\equiv} L_z^0 {-} L_z$. To make the numerics more tractable, as customary in the literature, we also assume a cutoff $2N{-}1$ for the total number of orbitals accessible to the particles. We note that, due to the additional step of finding the kernel of $\hat M_\mathrm{COM}$, the diagonalization scheme on the disk geometry is significantly less efficient compared to the sphere or torus. 

The presence of an edge can severely impact the bulk properties of small systems. Nevertheless, Fig.~\ref{fig:disk} shows that this is not the case here, and we can resolve the collective mode despite a fairly small number of particles. The softening of the Coulomb interaction by reducing $V_0$ has a similar effect as on the sphere and torus geometries: it makes the magnetoroton mode gapless in the long-wavelength limit, resulting in a nematic instability. In principle, by studying the momentum sectors $L_z {>} L_z^0$, the effect of nematic fluctuations on FQH edge states could also be explored in this setup, which we leave for future work. 

\begin{figure*}
    \centering
    \includegraphics[width=\linewidth]{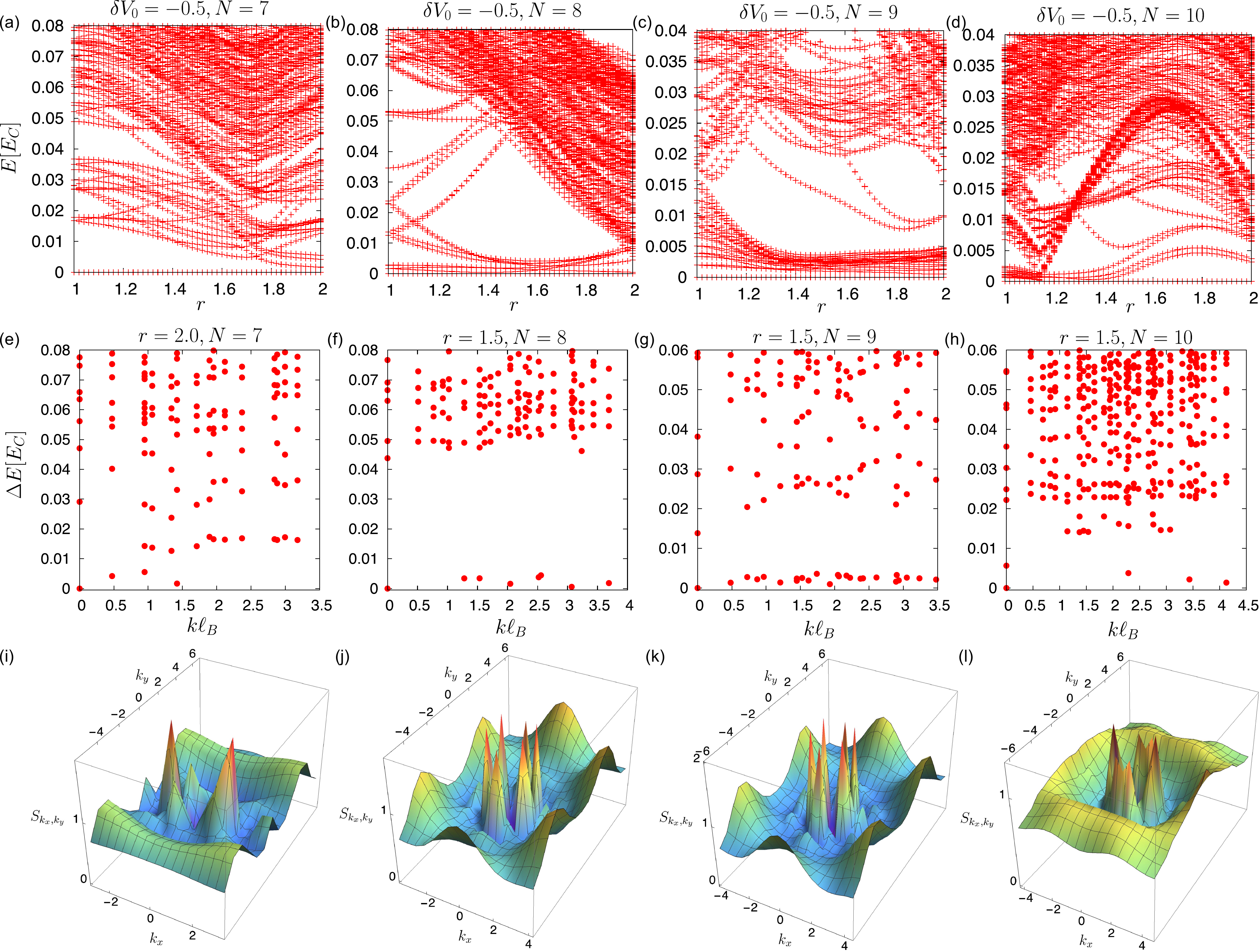}
    \caption{(a)-(d): Energy spectra for bosons at filling factor $\nu_b{=}1/2$ as a function of torus aspect ratio (for a rectangular unit cell). The interaction is Coulomb with $\delta V_0{=}{-}0.5$, i.e., past the nematic critical point. Large quasi-degeneracies are visible in the ground-state manifold, which is consistent with different types of CDW (see text). 
    (e)-(h): Momentum-resolved energy spectrum at illustrative aspect ratios $r{=}1.5$ (for $N{=}8,9,10$) and $r{=}2$ ($N{=}7$). The spectrum is plotted relative to the ground state in each case. 
    (i)-(l): Guiding center structure factor $S_{k_x,k_y}$ for one of the ground states in panels (e)-(h). In all the cases, we chose the ground state in momentum sector $(k_x{=}0,k_y{=}0)$. The distribution of momenta, belonging to the ground state multiplet, across the Brillouin zone and the profile of $S_{k_x,k_y}$ suggest that the system realizes a stripe at $N{=}7$, while it is closer to a bubble phase at $N{=}8,9,10$.
    }
    \label{fig:torus_cdw}
\end{figure*}

\section{Computation of the Hall viscosity}\label{app: viscosity}

To compute the Hall viscosity, we consider a torus defined by two sides $L_1$ and $L_2{=}L_1\tau$ in the complex plane, where $\tau{=}\tau_1{+}i\tau_2$ is the modular parameter. We assume $L_1$ to be real and $\tau_2{>}0$. The total area is $A{=}L_1^2\tau_2$. Ref.~\cite{Avron95} has shown that the antisymmetric part of viscosity for an FQH fluid, which was named the Hall viscosity $\eta^A$, can be derived from the Berry curvature $\mathcal{F}_{\tau_1,\tau_2}$ in $\tau$ space as 
\begin{equation}
\label{berry-curv}
\eta^A=-\frac{\hbar \tau_2^2  }{ A}\mathcal{F}_{\tau_1,\tau_2},     
\end{equation}
where 
\begin{equation}
\label{BC}
\mathcal{F}_{\tau_1,\tau_2}=-2{\rm Im}\bigg\langle \frac{\partial \Psi}{ \partial \tau_1} \bigg|
\frac{\partial \Psi }{ \partial \tau_2}\bigg\rangle  
\end{equation}
and $|\Psi\rangle$ is the incompressible FQH ground state in the torus geometry. 

The Berry curvature for a single particle in the $n$th Landau level (with $n{=}0$ for LLL) is \cite{Levay95}
\begin{equation}\label{eq: Levay}
\mathcal{F}_{\tau_1,\tau_2}=-\frac{1}{2}\left(n+\frac{1}{2}\right)\frac{1}{\tau_2^2}. 
\end{equation}
The Hall viscosity for an integer number of $n$ filled LLs can then be straightforwardly derived by adding the contributions of all single particles (which is valid for a Slater determinant state) to give
\begin{equation}
\eta^A_n=n\frac{\hbar}{4}\frac{N}{A}.     
\end{equation}
For FQH states, the Hall viscosity can be computed from the overlaps of the ground states with deformations of the geometry, i.e., the values of $\tau$. Hall viscosity is expected to be quantized within a FQH phase as \cite{Read09}
\begin{equation}
\label{hall visc}
\eta^A=\mathcal{S}\frac{\hbar}{4}\frac{N}{A},    
\end{equation}
where $\mathcal{S}$ is the so-called ``shift", i.e., the offset of flux quanta needed to form a ground state on a sphere:
 \begin{equation}
 \mathcal{S}=\frac{N}{\nu}-N_\phi     
 \end{equation}
where $N$ is the electron number, $N_\phi$ is the flux quanta number, and $\nu$ is the filling factor. The relation Eq.~\eqref{hall visc} has been confirmed through various calculations \cite{Read11, Tokatly09, Fremling14, Pu20}. 

The Berry curvature defined in Eq.~\eqref{BC} can be calculated if the exact ground state is known. To simplify the computation, we first obtain the ground states at four different points $\tau{=}(\tau_1{\pm}\delta_1,\tau_2{\pm}\delta_2)$ using exact diagonalization. We refer to these four points $\tau(m{=}1,2,3,4)$ in counter-clockwise order around $\tau$. Suppose the exact ground state eigenvectors are $|u_m\rangle$, $m{=}1,2,3,4$, respectively. The Berry curvature can be evaluated as 
\begin{eqnarray}
\label{BC2}
\nonumber \mathcal{F}_{\tau_1,\tau_2} = &-& \frac{1}{2\delta_1\delta_2}{\rm Im}\left[\ln\left( \langle u_1| u_2\rangle \langle u_2|u_3\rangle \langle u_3|u_4\rangle \langle u_4 | u_1\rangle \right)\right]\\
&-& \frac{N}{4\tau_2^2}.    
\end{eqnarray}
This is essentially a Wilson loop calculation of the Berry phase, with the advantage that we can avoid aligning the gauge choice at different $\tau$. The second term in Eq.~\eqref{BC2} accounts for the contribution from the change of the Slater determinant basis with a variation of $\tau$, which is a constant and equivalent to the Berry curvature of the filled LLL, directly following from Eq.~\eqref{eq: Levay}. In the calculation presented in the main text, we chose $\tau{=}i$, $\delta_1{=}0.001\pi$ and $\delta_2{=}0.001$ (thus, the variation of angle is 0.001 in units of $\pi$).

\begin{figure}
    \centering
    \includegraphics[width=\linewidth]{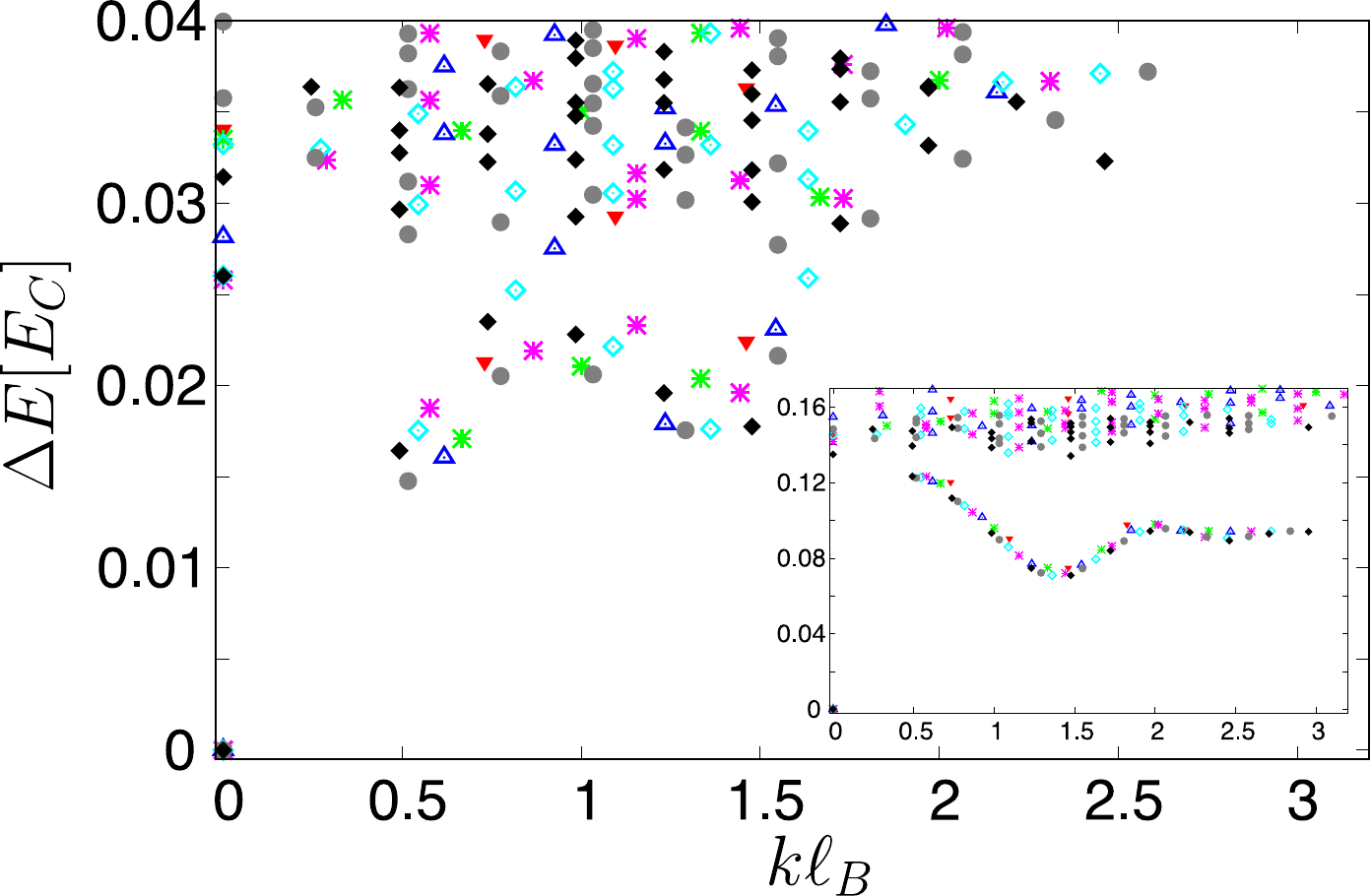}
    \caption{Energy spectrum for fermions at $\nu{=}1/3$ in the LLL interacting via Coulomb interaction with added $\delta V_1 {=} {-}0.09$ pseudopotential, which pushes the system to the vicinity of the critical point. Data is for system sizes $N{=}6{-}12$. Inset shows the spectrum for the pure Coulomb potential.}
    \label{fig: fermions_spectrum}
\end{figure}

\begin{figure}
    \centering
    \includegraphics[width=0.95\linewidth]{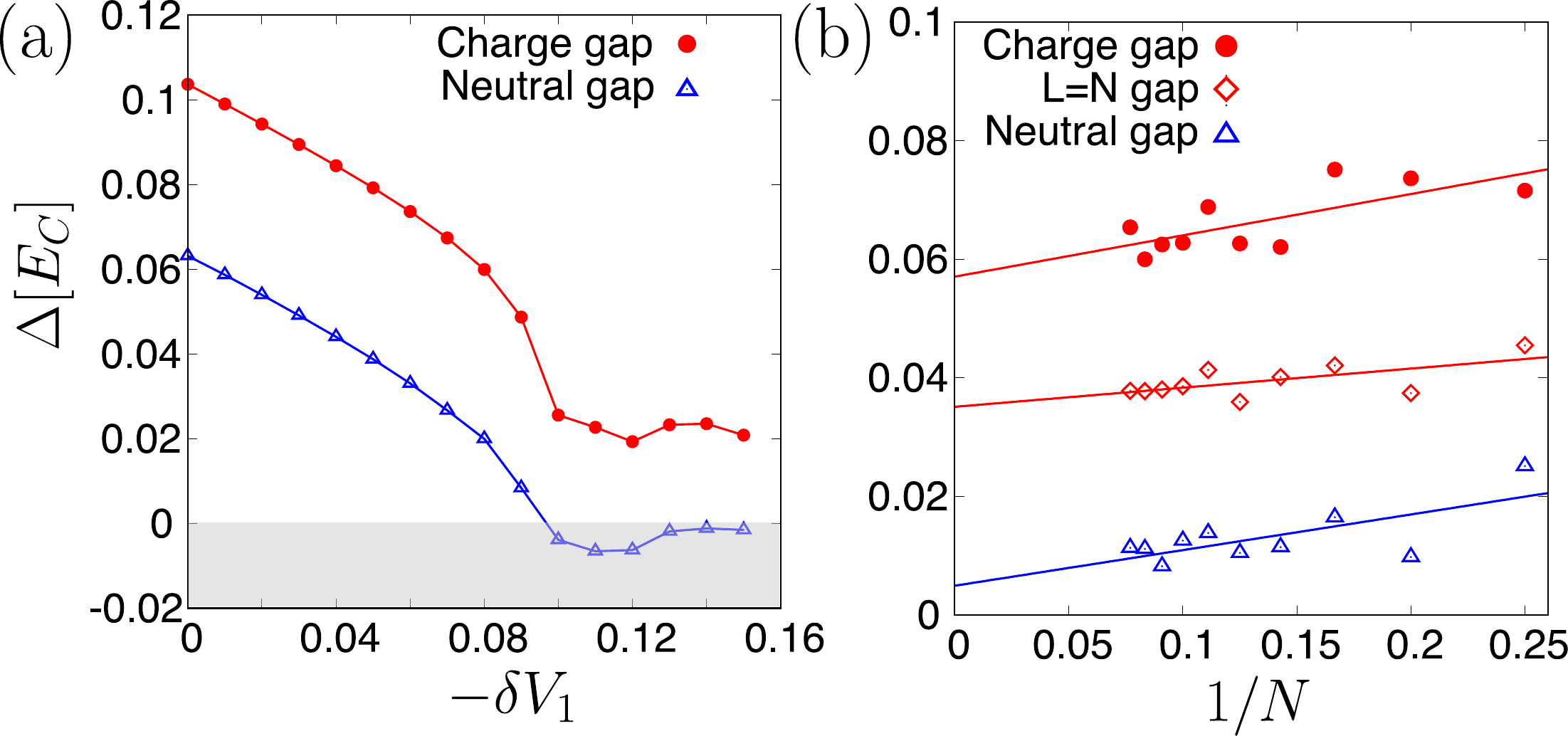}
    \caption{(a) The LLL Coulomb neutral and charge gaps as a function of $\delta V_1$ for the fermionic $\nu{=}1/3$ state. (b) The gaps as a function of $1/N$ for $\delta V_1{=}{-}0.095$ for the fermionic $\nu{=}1/3$ state. The finite-size effects are stronger in this case compared to bosons in the main text, 
    as can be seen from the difference of the extrapolated charge and $L{=}N$ gaps. Nevertheless, the lower value of these gaps is still by a factor of 3 larger than the neutral gap, which extrapolates close to zero.}
    \label{fig: fermions_gaps}
\end{figure}

\section{Charge density wave} \label{app: cdw}

In the main text, we have stated, without proof, that a CDW phase is expected to occur when the interaction becomes nearly hollow core. Consequently, a melted, translationally-invariant, form of such a CDW could explain the region of our phase diagram that was identified with the FQH nematic. Here we investigate in more detail the signatures of a CDW phase on the torus geometry, fixing the rectangular unit cell and varying its aspect ratio. It has been shown that a careful analysis of the energy spectrum and properties of the ground state under such deformations of the torus can be used to identify different types of CDW ordering, including stripe phases at half filling of higher LLs~\cite{Rezayi99Stripe}, electron bubble phases~\cite{Haldane00Bubble}, and Wigner crystals at low-filling factors~\cite{Yang01WC}, which can be viewed as a special case of one-electron bubbles.  
In Fig.~\ref{fig:torus_cdw} we perform a similar analysis for the present case of bosons at $\nu_b{=}1/2$ with the Coulomb interaction softened by $\delta V_0{=}{-}0.5$. This value of $\delta V_0$ is sufficiently large to push the system across the nematic critical point.

The low-lying energy spectra on the torus, Fig.~\ref{fig:torus_cdw}(a)-(d), display significant differences across the studied system sizes $N{\leq}10$. Nevertheless, as the torus aspect ratio is varied in the vicinity of the square unit cell, in all system sizes there appears a quasi-degenerate ground state manifold. The observed degeneracy is much larger than the ones found in incompressible FQH states~\cite{Haldane85b}. Instead, the momenta of degenerate states form a one- or two-dimensional lattice in reciprocal space. The quasi-degeneracy can be seen clearly in Figs.~\ref{fig:torus_cdw}(e)-(h), which show the momentum-resolved spectra for a fixed aspect ratio that is approximately optimal for a CDW in a given finite-size system, as seen in in Figs.~\ref{fig:torus_cdw}(a)-(d). For example, for $N{=}7$ bosons, we took the aspect ratio $r{=}2$ and plotted the spectrum in Fig.~\ref{fig:torus_cdw}(e). The momenta of the four lowest-lying states in Fig.~\ref{fig:torus_cdw}(e) form a one-dimensional array, similar to Ref.~\cite{Rezayi99Stripe}. Thus, we identify this ground state as a stripe. This identification is further corroborated by evaluating the guiding-center structure factor $S_{k_x,k_y}$~\cite{Prange87}, which is shown in Fig.~\ref{fig:torus_cdw}(i) for one of the quasi-degenerate ground states with momentum $(k_x{=}0,k_y{=}0)$. The structure factor exhibits sharp peaks that are colinear and separated by a multiple of a fundamental vector $\mathbf{k}^*$, indicating CDW ordering along one spatial direction.

In larger system sizes, $N{=}8,9$ in Figs.~\ref{fig:torus_cdw}(f)-(g), the ground state degeneracy is larger than expected for a stripe. For example, at $N{=}8$ and aspect ratio $r{=}1.5$, the number of quasi-degenerate ground states in Fig.~\ref{fig:torus_cdw}(f) is 16 and they form a two-dimensional lattice of 2-electron bubbles~\cite{Haldane00Bubble}. The corresponding structure factor of the ground state in  $(k_x{=}0,k_y{=}0)$ momentum sector is shown in Fig.~\ref{fig:torus_cdw}(j). The structure factor also exhibits sharp peaks but now they occur in both spatial directions. The overall shape of $S_{k_x,k_y}$ is similar to the examples of bubble phases in the third LL given in Ref.~\cite{Haldane00Bubble}. 

Note that system size $N{=}10$, shown in Fig.~\ref{fig:torus_cdw}(h), differs somewhat from $N{=}8,9$ by a visibly smaller ground-state degeneracy. At the same time, the structure factor shown in Fig.~\ref{fig:torus_cdw}(l) is broadly similar to those at $N{=}8,9$. Thus, it is possible that $N{=}10$ also realizes a bubble phase, but the splitting between the ground states is much larger than in  $N{=}8,9$. However, it is not clear at present why the splitting should be larger in this case.

In summary, Fig.~\ref{fig:torus_cdw} demonstrates that bosons at $\nu_b{=}1/2$ display signatures of CDW ordering once $\delta V_0$ is sufficiently reduced to cross the nematic critical point. While the nature of the CDW is hard to pinpoint due to strong finite-size fluctuations, all studied systems display large ground state quasi-degeneracy and sharp peaks in the structure factor, consistent with a CDW.

\begin{figure}
    \centering
    \includegraphics[width=\linewidth]{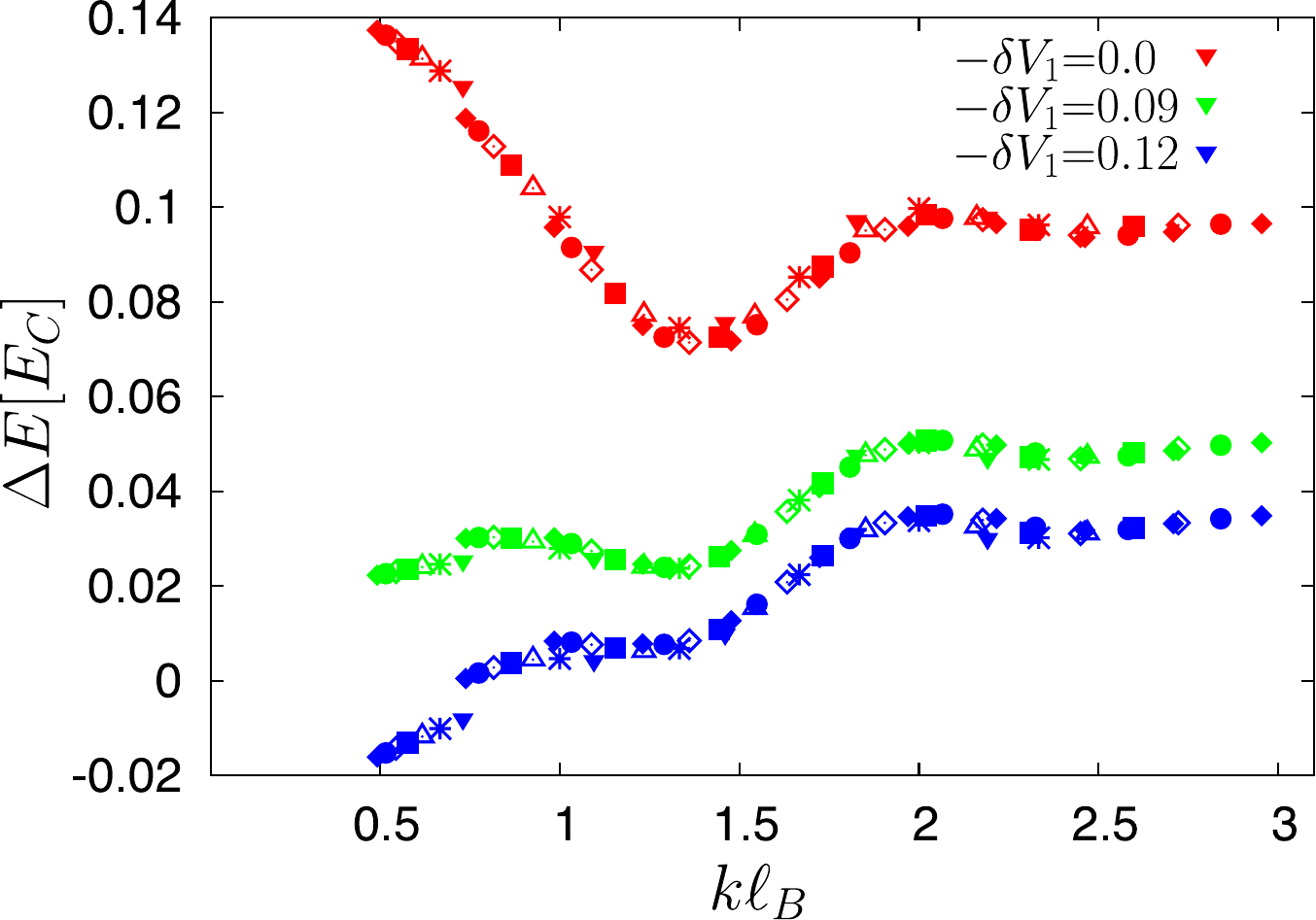}
    \caption{Variational energies for the collective mode states for fermions at $\nu{=}1/3$, approximated by Jack polynomials from Ref.~\cite{Yang12b}. Data is for system sizes $N=6-12$ and three values of $\delta V_1$ specified in the legend. We observe similar softening of the collective mode in the long-wavelength limit like in the bosonic case in the main text. 
    }
    \label{fig: fermions_jack}
\end{figure}

\section{Fermionic Laughlin state}\label{app: fermions}

In this section, we demonstrate that our results for bosons at $\nu_b{=}1/2$ in the main text are qualitatively similar to results for fermions at $\nu{=}1/3$, which is more experimentally relevant for solid-state materials. As will become apparent from the results below, the main difference is the stronger finite-size effects for fermions. This can be understood from the CF picture: the effective magnetic length for CFs is $\ell_B^{*}{=}\sqrt{2}\ell_B$ at $\nu_{b}{=}1/2$ while it is $\ell_B^{*}{=}\sqrt{3}\ell_B$ at $\nu{=}1/3$. Thus, the quasiparticles have larger sizes in the $1/3$ state compared to $1/2$, requiring larger systems to minimize the finite-size effects.

In Fig.~\ref{fig: fermions_spectrum} we show the exact spectrum of the LLL-projected Coulomb interaction for fermions at $\nu{=}1/3$ as the interaction is softened by $\delta V_1{=} {-}0.09$. By contrast, the inset to Fig.~\ref{fig: fermions_spectrum} shows the spectrum of the pure Coulomb interaction. Compared to the bosonic spectra in the main text 
the key difference is a more pronounced magnetoroton minimum for the case of pure Coulomb interaction in the fermionic case. Nevertheless, the softening of the interaction similarly leads to the collective mode going soft in the $k{\to} 0$ limit, though stronger finite-size effects are present in the fermionic case as seen in the spreading of the data points in Fig.~\ref{fig: fermions_spectrum}.

\begin{figure}
    \centering
    \includegraphics[width=\linewidth]{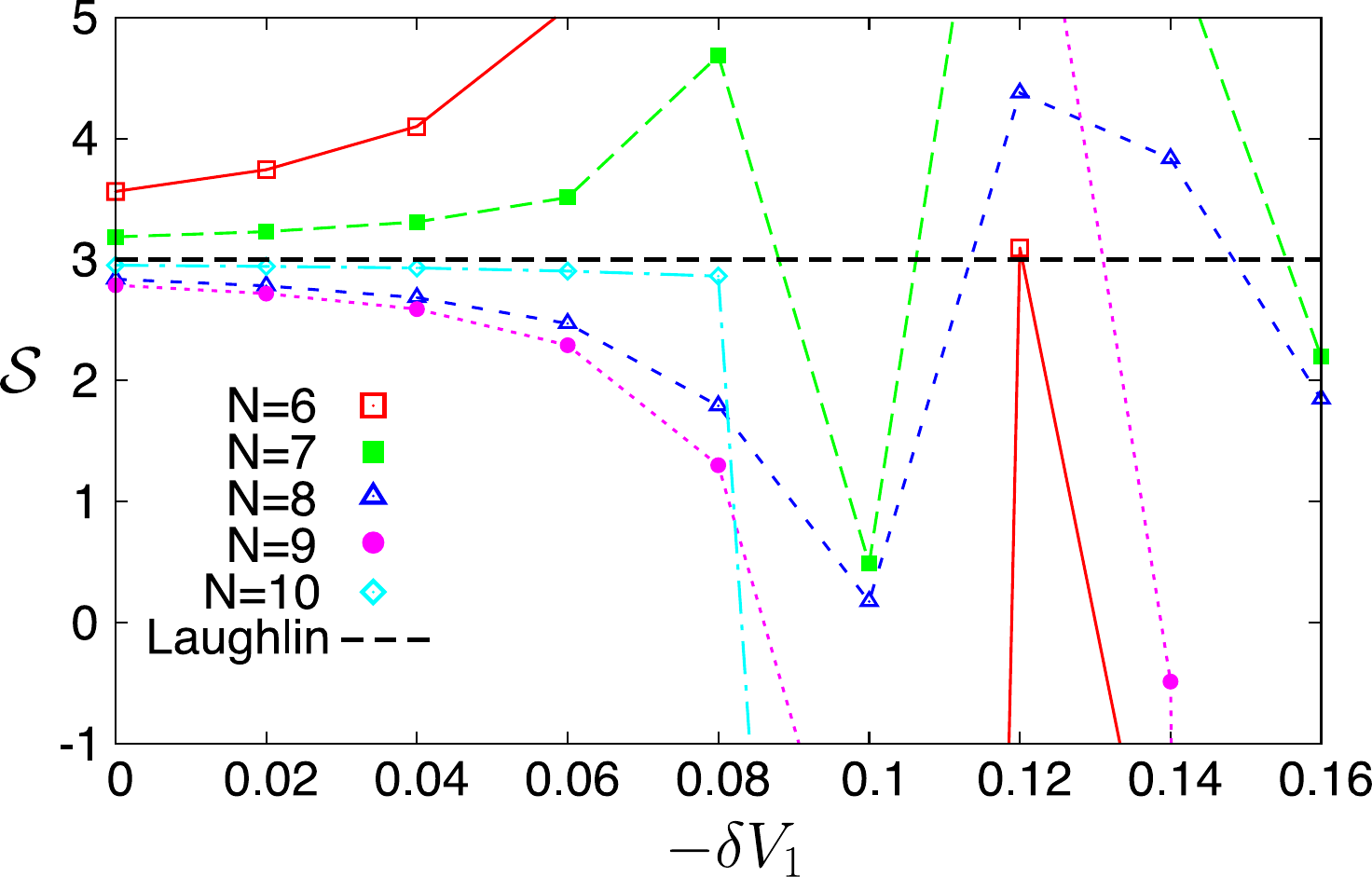}
    \caption{
    The shift extracted from
the Hall viscosity $\eta^A$ for fermions at $\nu{=}1/3$ on the torus in the vicinity of the square
aspect ratio. In the gapped FQH phase, the shift is quantized
to the value expected in the Laughlin state, $\mathcal{S}{=}3$, similar to the bosonic case in the main text. As the FQHN critical point is approached, large fluctuations in $\mathcal{S}$ appear and its value is no
longer quantized.}
    \label{fig: fermions_viscosity}
\end{figure}

The behavior of the extrapolated charge and neutral gaps for fermions at $\nu{=}1/3$ is shown in Fig.~\ref{fig: fermions_gaps}(a)-(b), respectively. The qualitative behavior of the gaps is similar to the bosonic case in the main text. 
The neutral gap closes at about $\delta V_1 {\approx} {-}0.095$. At the same point, the charge gap remains open and hovers around  $0.02E_C$ upon further reduction of $V_1$. Once again, the stronger finite-size effects compared to the bosonic case are also present in the gap extrapolations in Fig.~\ref{fig: fermions_gaps}(b). In particular, the charge gap and $L{=}N$ gap do not extrapolate to the same value as they do for the bosonic case in the main text. Nevertheless, both gaps remain by at least a factor of 3 larger than the extrapolated neutral gap at the FQHN transition ($\delta V_1 {=} {-}0.095$).

Finally, the behavior of the Hall viscosity for fermions at $\nu{=}1/3$ in Fig.~\ref{fig: fermions_viscosity} largely mirrors that of the bosonic case. 
Here the stronger finite-size effect is particularly clear, even deep within the Laughlin phase ($\delta V_1 {\geq} {-}0.1$). In fact, even at $\delta V_1 {=} 0$, the extracted shift shows a visible deviation from the Laughlin $\mathcal{S}{=}3$ in smaller system sizes $N{\leq} 9$, for which it was essentially converged in the bosonic case.

\end{document}